\documentclass[twocolumn]{svjour3}
\usepackage{amsmath}
\usepackage{amsfonts}
\usepackage{mathptmx}
\usepackage{refcount}
\usepackage[breaklinks=true]{hyperref}
\usepackage{natbib}
\usepackage{graphicx,fleqn}
\newcommand*{\ol}[1]   {\overline{#1}}
\newcommand*{\p}       {\partial}
\newcommand*{\Exp}[1]  {\mathrm{e}^{#1}}
\DeclareSymbolFont{AMSa}{U}{txsya}{m}{n}
\DeclareMathSymbol{\lesssim}{\mathrel}{AMSa}{46}
\DeclareMathSymbol{\gtrsim}{\mathrel}{AMSa}{38}
\journalname{Computational Astrophysics and Cosmology}
\title{A fast multipole method for stellar dynamics}
\titlerunning{FMM for stellar dynamics}
\author{Walter Dehnen}
\institute{Department for Physics \& Astronomy,
  University of Leicester,
  Leicester LE1 7RH, United Kingdom.
  \email{walter.dehnen@le.ac.uk}
}
\date{\today}
\begin{document}
\maketitle
\label{firstpage}
\begin{abstract}
  The approximate computation of all gravitational forces between $N$
  interacting particles via the fast multipole method (FMM) can be made as
  accurate as direct summation, but requires less than $\mathcal{O}(N)$
  operations. FMM groups particles into spatially bounded cells and uses
  cell-cell interactions to approximate the force at \emph{any} position
  within the sink cell by a Taylor expansion obtained from the multipole
  expansion of the source cell. By employing a novel estimate for the errors
  incurred in this process, I minimise the computational effort required for a
  given accuracy and obtain a well-behaved distribution of force errors.  For
  relative force errors of $\sim10^{-7}$, the computational costs exhibit an
  empirical scaling of $\propto N^{0.87}$. My implementation (running on a 16
  core node) out-performs a GPU-based direct summation with comparable force
  errors for $N\gtrsim10^5$.
\end{abstract}
\section{Introduction}
\label{sec:intro}
The computation of the mutual gravitational forces at every time step dominates
the computational costs of all $N$-body simulations. When simulating
collisionless stellar dynamics, the $N$-body model is merely a Monte-Carlo
representation of a smooth phase-space distribution and the $N$-body force is
only ever an \emph{estimate} for the smooth force field of the continuous system
modelled \citep[see also][]{DehnenRead2011}. In particular, the $N$-body force
unavoidably carries an \emph{estimation error}. This motivates the use of
\emph{approximate} methods for computing the $N$-body force, such as the
\cite{BarnesHut1986} tree code, as long as the approximation errors are small
compared to the estimation errors.

$N$-body simulations of collisional stellar dynamics are of a completely
different nature. Here, the particles simulate individual stars and the $N$-body
force carries no estimation error. Consequently, the (negative) gravitational
potential
\begin{equation} \label{eq:P2P}
  \Psi(\vec{x}_b) = \sum_{a\neq b} G\mu_a\,\psi(\vec{x}_b-\vec{x}_a)
  \quad\text{with}\quad
  \psi(\vec{r}) = |\vec{r}|^{-1}
\end{equation}
and its derivative, the acceleration, must be calculated with high accuracy.
This is typically achieved by \emph{direct summation}, when
equation~(\ref{eq:P2P}) is translated into computer code and the only errors are
owed to finite computational precision.

This computation incurs a cost of $\mathcal{O}(N)$ for a single particle and
thus $\mathcal{O}(N^2)$ per unit time for running a full simulation. As a
consequence, realistic simulations with $N\sim10^{6-7}$ for globular clusters
and galactic centres are still very challenging and large parameter studies
impossible. Measures employed to ameliorate this situation include the usage of
powerful special-purpose hardware devices \citep{MakinoTaiji1998} or graphical
processing units \citep[GPUs,][]{GaburovHarfstPortegiesZwart2009}, as well as
separating the highly fluctuating forces due to close neighbours, in order to
reduce the frequency of expensive far-field force computations
\citep{AhmadCohen1973}.

While these measures substantially reduce the effective costs, the complexity
of $N^2$ remains. The alternative of using approximate methods also for
collisional stellar dynamics is so far untested. The requirements for such a
method differ from that in collisionless $N$-body methods in two important
aspects: (i) there is no gravitational softening and (ii) to preserve the
validity of the $N$-body model, the approximation errors must be much smaller
than what is common in collisionless $N$-body simulations.

A straightforward approach is to use the tree code with a small opening angle
and/or high expansion order, resulting in a scheme with $\mathcal{O}(N\ln N)$
costs. A more efficient approach is to use the \emph{fast multipole method}
\citep*[FMM;][]{GreengardRokhlin1987, ChengGreengardRokhlin1999}, which has
costs of only $\mathcal{O}(N)$. An initial attempt by
\cite{CapuzzoDolcettaMiocchi1998} to port this technique from its original
realm of molecular dynamics to astrophysics failed to obtain better practical
efficiency than the tree code. However, when adapting the FMM to the
inhomogeneity of stellar systems and the low force accuracy required in
collisionless dynamics (by using a hierarchical tree data structure and a
flexible opening angle), it is substantially faster than the tree code
\citep{Dehnen2000:falcON, Dehnen2002}.

The critical question here is whether FMM can be tuned to be more efficient
than direct summation at force accuracies and particle numbers required by
collisional $N$-body techniques. The goal of this study is to address this
question by tuning FMM for the application to collisional $N$-body
simulations, investigating the resulting dependence of computational costs and
numerical accuracy on the various numerical parameters, and assessing its
practical efficiency.

This paper is organised as follows. In Section~\ref{sec:prelim} and
Appendix~\ref{app:fmm}, the mathematical (and algorithmic) foundations of FMM
are derived and laid down. Section~\ref{sec:quant:error} (and
Appendix~\ref{app:error}) introduces and motivates my approach for quantifying
the resulting acceleration errors; Section~\ref{sec:errors} provides useful
estimates for the errors of individual FMM interactions; Section~\ref{sec:mac}
deals with optimising the multi\-pole-accep\-tance criterion; and in
Section~\ref{sec:opt} the method is tuned to obtain a force accuracy target with
minimal computational effort. Finally, in Section~\ref{sec:beyond} possible
extensions and applications are discussed, and Section~\ref{sec:conclude}
concludes.

\section{FMM basics}
\label{sec:prelim}
The tree code approximates the sum~(\ref{eq:P2P}) by first dividing source
particles $a$ into groups bounded by geometric cells, each of which is
well-separated from the sink position $\vec{x}_b$, and then computing the forces
of each source cell from their multipole moments. This corresponds to Taylor
expanding the Greens function $\psi(\vec{x}_b-\vec{x}_a)$ about the distance to
an appropriate centre $\vec{z}$ of each source cell.

The essence of the fast multipole method is to Taylor expand the Greens function
not only at the source positions $\vec{x}_a$, but also at the sink positions
$\vec{x}_b$. This latter amounts to approximating (a contribution to) the
gravitational field within each sink cell by its local Taylor expansion about
some appropriate potential expansion centre $\vec{s}$. Obviously, this approach
is beneficial only if the forces for a large fraction of the sinks within a cell
are to be computed simultaneously.
 
\subsection{Mathematical background}
\label{sec:prelim:math}
The FMM relations are most easily derived using Cartesian
coordinates. However, for Newtonian gravity, $\psi=|\vec{r}|^{-1}$, the
resulting relations are inefficient. Instead, exploiting that this Greens
function satisfies $\vec{\nabla}^2\psi=0$ for $\vec{r}\neq0$ naturally leads
to spherical harmonics. \cite{ChengGreengardRokhlin1999} have already given
(without derivation) the corresponding FMM relations, but in a form ill-suited
for computer code. In Appendix~\ref{app:fmm}, I derive equivalent but much
more compact and computationally convenient relations. These are summarised
here.

Let $\vec{r}=(x,y,z)$ with spherical polar coordinates $r,\theta,\phi$, then
\begin{subequations}
  \label{eqs:Ups:The}
  \begin{eqnarray}
    \label{eq:The}
    \Theta_n^m(\vec{r}) &=& (-1)^m \frac{(n-m)!}{r^{n+1}}
    P_n^{m}(\cos\theta)\,\Exp{im\phi},\\[0.5ex]
    \label{eq:Ups}
    \Upsilon_n^m(\vec{r}) &=& (-1)^m \frac{r^n}{(n+m)!}
    P_n^{m}(\cos\theta)\,\Exp{im\phi}
  \end{eqnarray}
\end{subequations}
with integer indices $0\le|m|\le n$ are (complex-valued) harmonic
functions, i.e.\ $\vec{\nabla}^2\Upsilon_n^m=0$ for all $\vec{r}$ and
$\vec{\nabla}^2\Theta_n^m=0$ for all $\vec{r}\neq0$. The $\Upsilon_n^m$ are
homogeneous polynomials of total degree $n$ in $x$, $y$, and $z$ (they are
defined in Appendix~\ref{app:spherical:complex} without reference to polar
coordinates; see also Table~\ref{tab:The}). With these definitions, the FMM
relations for the computation of the potential due to all particles within
source cell $A$ and at any position $\vec{x}_b$ within sink cell $B$ are
\begin{subequations}
  \label{eqs:fmm}
  \begin{eqnarray}
    \label{eq:L2P:Psi}
    \Psi_{A\to B}(\vec{x}_b) 
    &=&
    \sum_{n=0}^p\,\sum_{m=-n}^n\Upsilon_n^{m\ast}(\vec{s}_{B}-\vec{x}_b)\,
    \mathcal{F}^m_n(\vec{s}_{B}) \,+\, \delta\Psi_{A\to B},
    \\[0.5ex]
    \label{eq:M2L}
    \mathcal{F}^{m}_{n}(\vec{s}_{B}) &=&
    \sum_{k=0}^{p-n}\sum_{l=-k}^{k}\mathcal{M}^{l\ast}_{k}(\vec{z}_{A})\,
    \Theta_{n+k}^{m+l}(\vec{s}_{B}-\vec{z}_{A}),
    \\[0.5ex]
    \label{eq:P2M}
    \mathcal{M}^m_{n}(\vec{z}_{A}) &=& \sum_{a\in A} \mu_a\,
    \Upsilon_n^m(\vec{x}_a-\vec{z}_{A}).
  \end{eqnarray}
  Here, $p$ is the expansion order and $\delta\Psi_{A\to B}$ the error of the
  approximated potential. This expansion converges with increasing $p$ if
  $\max_{a\in A}\{|\vec{x}_b-\vec{x}_a-\vec{r}|\} < |\vec{r}|$ with
  $\vec{r}\equiv\vec{s}_{B}-\vec{z}_{A}$.

  Other important relations are those for the \emph{multipoles}
  $\mathcal{M}_n^m$ with respect to another expansion centre
  \begin{equation} \label{eq:M2M}
    \mathcal{M}_n^m(\vec{z}^\prime) = \sum_{k=0}^n\;\sum_{l=-k}^k
    \;\Upsilon_k^{l}(\vec{z}-\vec{z}^\prime)\,
    \mathcal{M}_{n-k}^{m-l}(\vec{z}),
  \end{equation}
  and for the \emph{field tensors} $\mathcal{F}_n^m$ of the local
  expansion~(\ref{eq:L2P:Psi}) with respect to another expansion centre
  \begin{equation} \label{eq:L2L}
    \mathcal{F}_n^{m}(\vec{s}^\prime) = \sum_{k=0}^{p-n} \sum_{l=-k}^k\,
    \Upsilon_k^{l\ast}(\vec{s}-\vec{s}^\prime)\,
    \mathcal{F}_{n+k}^{m+l}(\vec{s}).
  \end{equation}
  Moreover, the computation of the acceleration $\vec{a}$ from the local
  expansion (\ref{eq:L2P:Psi}) requires
  \begin{equation} \label{eq:L2P}
    \Psi_n^m(\vec{x}_b)=
    \sum_{k=0}^{p-n}\,\sum_{l=-k}^k
    \Upsilon_k^{l\ast}(\vec{s}_{B}-\vec{x}_b)\,
    \mathcal{F}_{n+k}^{m+l}(\vec{s}_{B})
    + \delta\Psi_n^m,
  \end{equation}
  when $\vec{a}=\vec{\nabla}\Psi_0^0=-(\Re\{\Psi_1^1\},
  \Im\{\Psi_1^1\},\Psi_1^0)$. Finally, the gravity generated from a source
  distribution with given multipoles is given by
  \begin{equation} \label{eq:M2P}
    \Psi_n^m(\vec{x}_b) =
    \sum_{k=0}^{p-n}\sum_{l=-k}^{k}\mathcal{M}_k^{l\ast}(\vec{z}_{A})\,
    \Theta_{n+k}^{m+l}(\vec{x}_b-\vec{z}_{A})
    \,+\, \delta\Psi_n^m.
  \end{equation}
\end{subequations}
Relations (\ref{eq:M2L}), (\ref{eq:M2M}), and (\ref{eq:L2L}) are equivalent to
the much more complicated equations (17), (13), and (21) of
\citeauthor*{ChengGreengardRokhlin1999} (\citeyear{ChengGreengardRokhlin1999},
given without derivation)\footnote{\citeauthor{ChengGreengardRokhlin1999}'s
  expressions are quite cumbersome because they are given in terms of the
  surface spherical harmonics $Y_n^m$ in polar coordinates and because they
  contain phase-factors like $i^{|m|-m}$ owing to their unconventional
  definition for the $Y_n^m$ which implies $Y_n^{-m}=Y_n^{m\ast}$ instead of
  $Y_n^{-m}=(-1)^mY_n^{m\ast}$.}.

\begin{figure}
  \centerline{\resizebox{70mm}{!}{\includegraphics{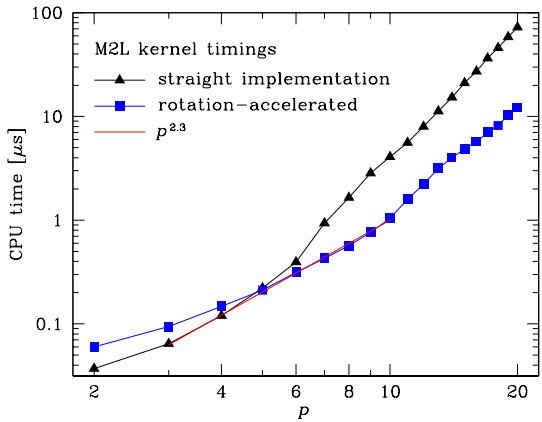}}}
  \caption{Timings for the M2L kernel (using double precision without
    vectorisation) as function of expansion order $p$. The rotation
    accelerated version of Appendix~\ref{app:fast:inter} is faster than the
    direct implementation of equation~(\ref{eq:M2L}) for $p\ge5$.
    \label{fig:interact}
  }
\end{figure}
\begin{figure*}
  \centerline{
    \resizebox{!}{25mm}{\includegraphics{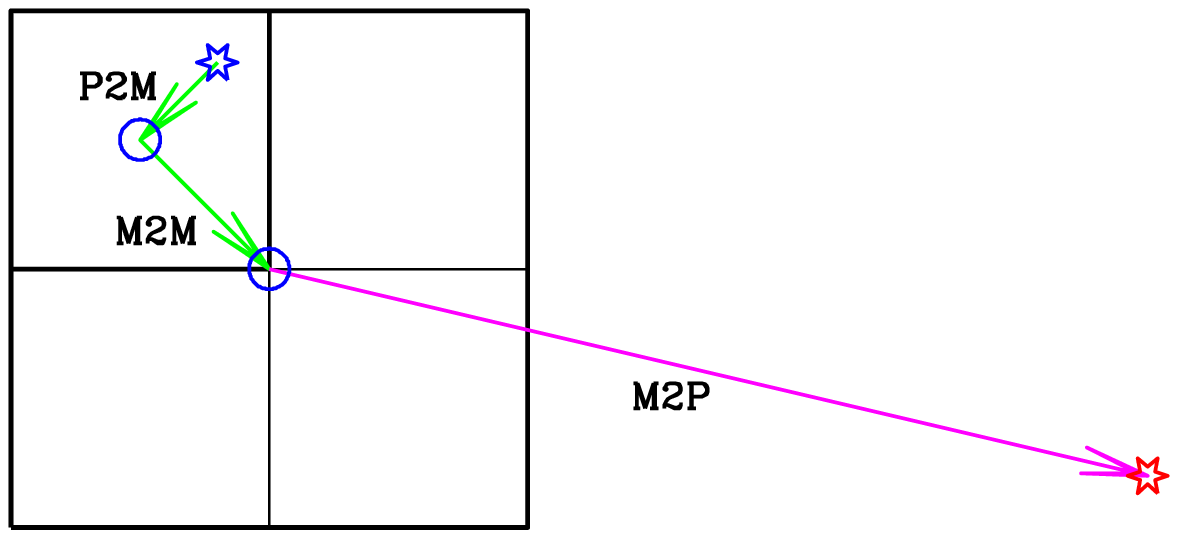}}\hspace*{30mm}
    \resizebox{!}{25mm}{\includegraphics{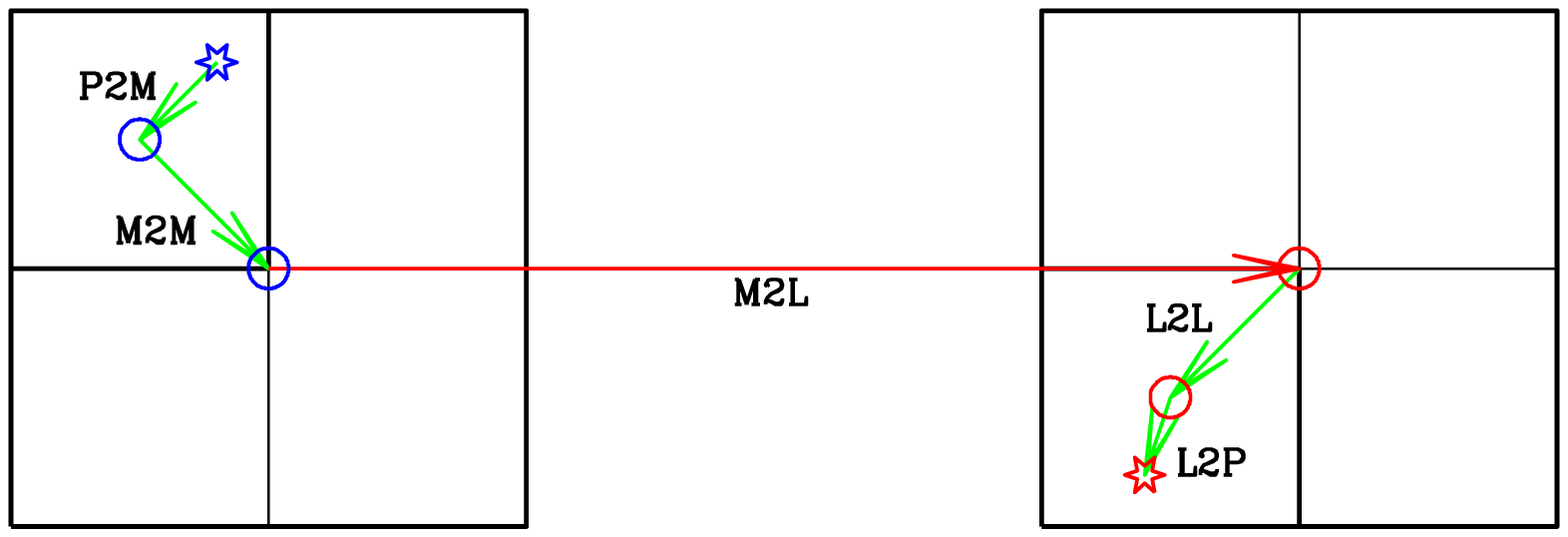}}
  }
  \caption{\textbf{Left}: the tree code approximates the force from a source
    particle (blue star) at a sink position (red star) using the P2M and M2M
    kernels (green arrows) to compute the multipoles at the cell centres (blue
    circles) followed by the M2P kernel (pink arrow). The P2M and M2M kernels
    are called once per source particle and cell, respectively, while the M2P
    kernel is called many times for each sink position. \textbf{Right}: FMM
    replaces the M2P kernel by the M2L kernel (red arrow) followed by the L2L
    and L2P kernels (green arrows). Again, the L2L and L2P kernels are called
    once per sink cell and particle, respectively, but a single M2L kernel
    replaces many M2P kernels of the tree code, because it accounts for
    \emph{all} sink positions within the sink cell.
    \label{fig:tree:fmm}
  }
\end{figure*}
There are $(p+1)^2$ independent real-valued numbers $\mathcal{F}_n^m$ (as well
as $\mathcal{M}_n^m$, see also Appendix~\ref{app:spherical:real}), and their
computation via equations~(\ref{eq:M2L}), (\ref{eq:M2M}), and (\ref{eq:L2L})
requires $\mathcal{O}(p^4)$ operations\footnote{\label{foot:O}Expressions like
  $\mathcal{O}(p^n)$ for the operation count relate to the asymptotic behaviour
  at large expansion orders $p$. While this is straightforward to specify, it is
  not necessarily very relevant, since in the range up to $p\sim10$, as required
  in practice, the actual costs usually grow more slowly than implied by the
  asymptotic behaviour (see Fig.~\ref{fig:interact} for a typical example) and
  because the numerical implementation may be data-dominated rather than
  computation dominated.}.
These operation counts can be reduced to $\mathcal{O}(p^3)$ by rotating
$\vec{r}$ into the $z$ direction (see Appendix~\ref{app:spherical:fast}).
Fig.~\ref{fig:interact} plots the time required per interaction computation as
function of expansion order $p$, showing an effective $p^{2.3}$ scaling of the
computational costs at $p\le10$, shallower than the $\mathcal{O}(p^3)$
asymptote.

\begin{table}
  \caption{The FMM \emph{kernels}. The tree code replaces direct summation
    (P2P) with P2M-M2M-M2P, while FMM uses P2M-M2M-M2L-L2L-L2P, see also
    Fig.~\ref{fig:tree:fmm}. The P2L kernel corresponds to
    equation~(\ref{eq:M2L}) with $\vec{z}_{A}\to\vec{x}_a$ and
    $\mathcal{M}_k^l\to m_a\delta_{k0}$, i.e.\ $\mathcal{F}_n^m(\vec{s}_{B})=
    m_a \Theta_n^m (\vec{s}_{B} - \vec{x}_a)$.
    \label{tab:kernels}
  }
  \begin{tabular}{lll}
    name & meaning & equation \\ \hline
    P2P & particle to particle & (\ref{eq:P2P}) \\
    P2M & particle to multipole & (\ref{eq:P2M}) \\
    M2M & multipole to multipole & (\ref{eq:M2M}) \\
    M2P & multipole to particle & (\ref{eq:M2P}) \\
    M2L & multipole to local expansion & (\ref{eq:M2L}) \\
    P2L & particle to local expansion & see table caption\\
    L2L & local expansion to local expansion & (\ref{eq:L2L}) \\
    L2P & local expansion to particle & (\ref{eq:L2P})
  \end{tabular}
\end{table}
\subsection{Algorithmic approach}
\subsubsection{The tree code: walking the tree}
Let us first consider the tree code, which also uses the multipole expansion but
is algorithmically simpler than FMM. The basic data structure is a hierarchical
tree of spatial cells, which are either cubic with eight daughters cells
(oct-tree) or cuboidal with two daughters (binary tree). In a first step, the
multipoles $\mathcal{M}_n^m$ have to be computed for each cell from those of
their daughter cells, using the M2M kernel (equation~\ref{eq:M2M}, see also
Table~\ref{tab:kernels}), or (in case of final cells) of their particles, using
the P2M kernel (equation~\ref{eq:P2M}).

Next, the force for each sink position is computed using a separate \emph{tree
  walk} starting with the root cell. The force generated by a cell $C$ is
computed via its multipole expansion, using the M2P kernel
(equation~\ref{eq:M2P}), if a \emph{multipole-acceptance criterion} is met,
i.e.\ if the cell is considered to be \emph{well-separated} from the sink
position.  Otherwise, the cell is opened: the force is computed as the sum of
the forces generated by the daughters cells (recursing if necessary). Thus, the
tree code replaces direct summation's P2P kernel with the P2M, M2M, and M2P
kernels, see the left panel of Fig.~\ref{fig:tree:fmm} for a schematic view.

\subsubsection{FMM: the dual tree walk}
\label{sec:algo:tree}
An adaptive FMM algorithm also uses a hierarchical tree data structure. As with
the tree code, the cell multipoles $\mathcal{M}_n^m$ have to be precomputed for
every cell in a first step.

Next, the forces for \emph{all} sink positions and generated by \emph{all}
source particles are approximated using a single \emph{dual tree walk}
\citep{Dehnen2002}. This algorithm considers cell$\,\to\,$cell interactions and
starts with the root$\,\to\,$root interaction. If the interacting cells are well
separated, the interaction is approximated using the M2L kernel
(equation~\ref{eq:M2L}), which computes and accumulates the local field tensors
$\mathcal{F}_n^m(\vec{s}_{B})$ for the expansion of gravity within the sink
cell $B$ and due to all sources within the source cell $A$ (in a mutual version
of the algorithm, the interactions $A\to B$ and $B\to A$ are considered
simultaneously). Otherwise, the interaction is split, typically into those
between the daughters of the larger of the two interacting cells with the
smaller.

Finally, the local field tensors $\mathcal{F}_n^m(\vec{s})$ are passed down the
tree using the L2L kernel, and the local expansions are evaluated at the sink
positions using the L2P kernel. Thus, the FMM replaces the M2P kernel of the
tree code with the M2L, L2L and L2P kernels, see also Fig.~\ref{fig:tree:fmm}.

Of course, in both tree code and FMM, direct summation (P2P kernel) is used
whenever computationally preferable, i.e.\ for interactions involving only a few
sources and sinks.

\section{Quantifying the approximation accuracy}
\label{sec:quant:error}
Before the method can optimised for accuracy, a sensible quantitative measure
for this accuracy is needed as well as an acceptable value for this measure.

With direct-summation, the accuracy is limited only by the finite precision of
computer arithmetic (round-off error). If double (64-bit) precision is not used
throughout, it is customary to use the conservation of the total energy for
quality control \citep[e.g.][]{GaburovHarfstPortegiesZwart2009}. However, as
shown in Appendix~\ref{app:error}, the relative energy error is much smaller
than the typical relative force error, simply because it is an average over many
force errors. Even worse, the computation of the total energy, required for
measuring its error, typically incurs a larger error. Thus, any measured
non-conservation of the total energy is dominated by measurement error rather
than true non-conservation due to acceleration errors.

With the tree code and FMM, the situation is subtly different, as discussed in
Appendix~\ref{app:error:approx}. Here, the measured non-conservation of energy
actually reflects the amplitude of the acceleration errors in an average sense.
However, an average measure for the effect of approximation errors cannot
reflect their effect on the correctness of the simulation. For example, a single
large force error has hardly any effect on the energy conservation but may
seriously affect the validity of the simulation.  While this latter goal is
difficult to quantify, it is certainly better to consider the whole distribution
of acceleration errors and pay particular attention to large-error outliers,
than merely monitor an average.

\subsection{Scaling acceleration errors}
\label{sec:error:scaled}
Obviously, the absolute errors $\delta
a=|\vec{a}_{\mathrm{computed}}-\vec{a}_{\mathrm{true}}|$ are not very useful by
themselves and must be normalised to be meaningful. One option is to divide
$\delta a$ by some mean field strength $\bar{a}$. While this makes sense for the
average particle, it fails for those in the outskirts of the stellar system,
where the field strength diminishes well below its mean.

To overcome such issues, a natural choice is the relative error $\delta a/a$.
However, this is still problematic in the centre of a stellar system, where
forces from the outward lying parts largely cancel. In such a situation, $a$ can
be small and hence the relative error large, even if each individual pair-wise
force has been computed with high accuracy. One option for avoiding this problem
is -- in analogy to the error estimate of numerical quadrature in case of an
integrand oscillating around zero -- to normalise
$\delta a$ with the sum
\begin{equation} \label{eq:f}
  f_b \equiv \sum_{a\neq b} \frac{G\mu_a}{|\vec{x}_a-\vec{x}_b|^2}
\end{equation}
of the absolute values of all pair-wise accelerations. In general $f_b\ge a_b$,
while in the outskirts of a stellar system $f\to a\approx GM/r^2$ such that the
\emph{scaled error} $\delta a/f$ approaches the relative error $\delta a/a$ as
desired. Conversely, in the centre $f\gg a$ (for a Plummer sphere, for example,
$f\to2GM/r_s^2$ as $r\to0$ in the continuum limit) and $\delta a/f$ behaves
sensibly if $a\to0$.

\begin{figure}
  \begin{center}
    \resizebox{\columnwidth}{!}{\includegraphics{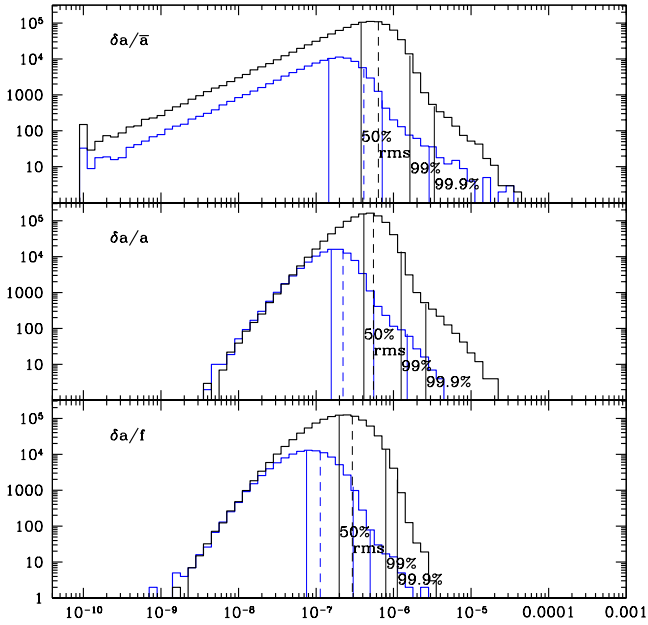}}
  \end{center}
  \caption{Distribution of acceleration errors $\delta a_i$ for $N=10^5$ (blue)
    and $N=10^6$ (black) particles drawn from a Plummer sphere when using the
    state-of-the-art GPU-based direct-summation library \textsf{sapporo}
    (version 1.6) as compared to direct summation in double (64bit)
    precision. The top, middle, and bottom panel refer to, respectively, the
    normalised (by the mean acceleration $\bar{a}$), relative, and scaled (by
    $f$ defined in equation~\ref{eq:f}) acceleration errors. The thin vertical
    lines indicate the rms error (dashed) as well as the median and the 99 and
    99.9 percentiles (solid). Bins are 0.1\,dex wide.
    \label{fig:da:gpu}
  }
\end{figure}
\subsection{The acceleration errors of direct summation}
\label{sec:error:dist:direct}
In order to assess the errors currently tolerated in collisional $N$-body
simulations, the GPU-based direct-summation library \textsf{sapporo}
\citep{GaburovHarfstPortegiesZwart2009} was applied to two sets of,
respectively, $N=10^5$ and $N=10^6$ equal-mass particles, drawn randomly from
a \cite{Plummer1911} sphere (without any outer
truncation). Fig.~\ref{fig:da:gpu} shows the resulting distributions of
acceleration errors as compared to direct summation in double (64bit)
precision. As expected, the typical relative (or scaled) error is
$\sim10^{-7}$, comparable to the relative round-off error of single-precision
floating-point arithmetic. However, there is a clear tail of large relative
errors (middle panel). This is due to particles at small radii, whose
acceleration is small, because the pair-wise forces with other particles
mostly cancel out, while the (round-off) errors accumulate.

There is a significant increase in the error amplitude with particle number $N$:
the errors for $N=10^6$ are on average $\sim\sqrt{10}$ larger than for
$N=10^5$. This worrying property suggests that the fidelity of simulations using
\textsf{sapporo} diminishes with $N$, implying that using this library with
$N\gtrsim10^7$ is not advisable.

From this exercise I conclude that in practice relative (or scaled)
acceleration errors with an rms value of a few $10^{-7}$ and maximum $\sim10$
times larger are accepted in $N$-body simulations of collisional stellar
dynamics.

\section{Assessing the approximation errors}
\label{sec:errors}
In order to optimise any implementation of FMM for high accuracy and low
computational costs, a good understanding of and accurate estimates for the
errors incurred by each individual FMM interaction are required. To this end, I
now perform some numerical experiments.

I create a Plummer sphere of $N=10^6$ particles and build an oct-tree. For each
cell, the centre $\vec{z}_{\mathrm{ses}}$ of the smallest enclosing sphere for
all its particles is found (see Section~\ref{sec:mac:z}). I use
$\vec{z}=\vec{s}=\vec{z}_{\mathrm{ses}}$ for each cell and pre-compute the
cells' multipole moments $\mathcal{M}_n^m(\vec{z})$.  Finally, the dual tree
walk is performed using the multipole-acceptance criterion
\begin{equation} \label{eq:mac:theta}
  \theta < \theta_{\mathrm{crit}}
\end{equation}
with the \emph{opening angle} 
\begin{equation} \label{eq:theta}
  \theta \equiv (\rho_{\vec{z},A}^{}+\rho_{\vec{s},B}^{})/r, \quad
  r\equiv|\vec{s}_{B}-\vec{z}_{A}|.
\end{equation}
Here, for each cell $C$
\begin{eqnarray}
  \rho_{\vec{z},C}^{} &\equiv&
  \max_{\mathrm{sources\,} a\in C} \{|\vec{x}_a-\vec{z}_C|\}
  \quad\text{and}\\
  \rho_{\vec{s},C}^{} &\equiv&
  \max_{\mathrm{sinks\,} b\in C}\{|\vec{x}_b-\vec{s}_C|\}
\end{eqnarray}
are (approximations for) the radii of the smallest spheres centred on $\vec{z}$
and $\vec{s}$ and containing all sources and sinks, respectively.  In the
experiments of this section $\rho_{\vec{z}}=\rho_{\vec{s}}$ for each cell,
because $\vec{z}=\vec{s}$ and because all particles are source and sink
simultaneously, but in general $\rho_{\vec{z}}$ and $\rho_{\vec{s}}$ may
differ.

\begin{figure*}
  \begin{center}
    \resizebox{78mm}{!}{\includegraphics{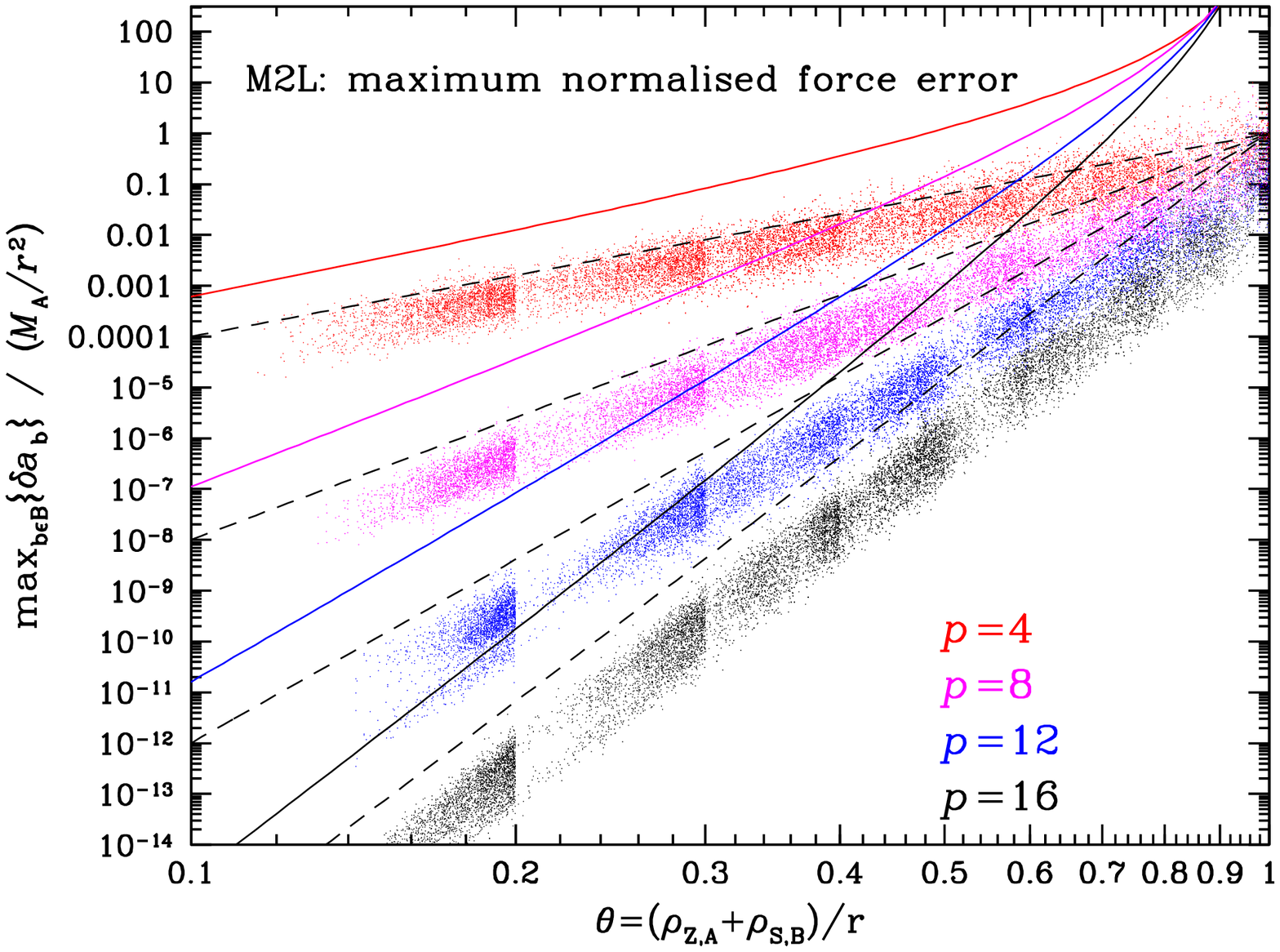}}\hspace{15mm}
    \resizebox{68mm}{!}{\includegraphics{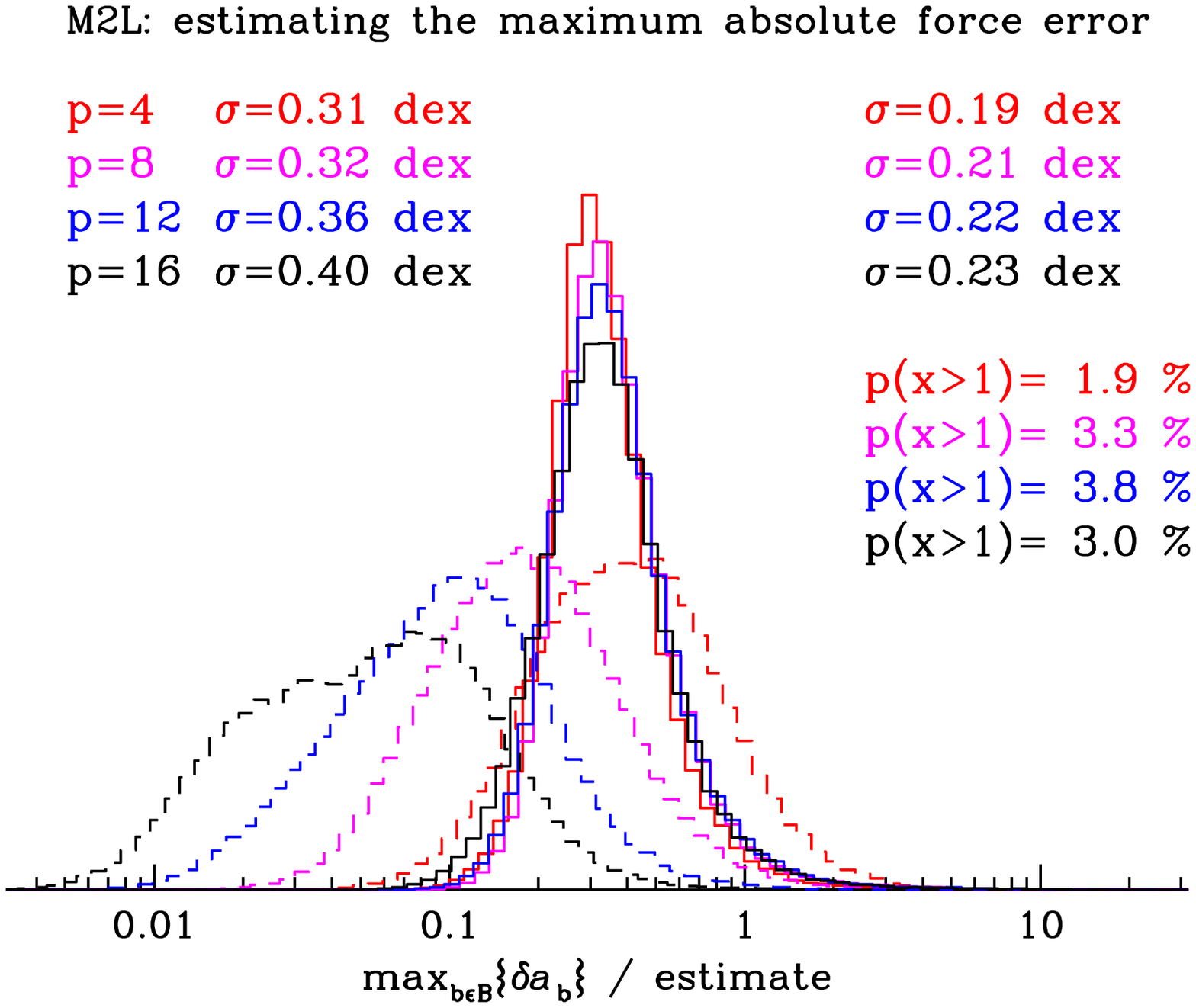}}\\[3mm]
    \resizebox{78mm}{!}{\includegraphics{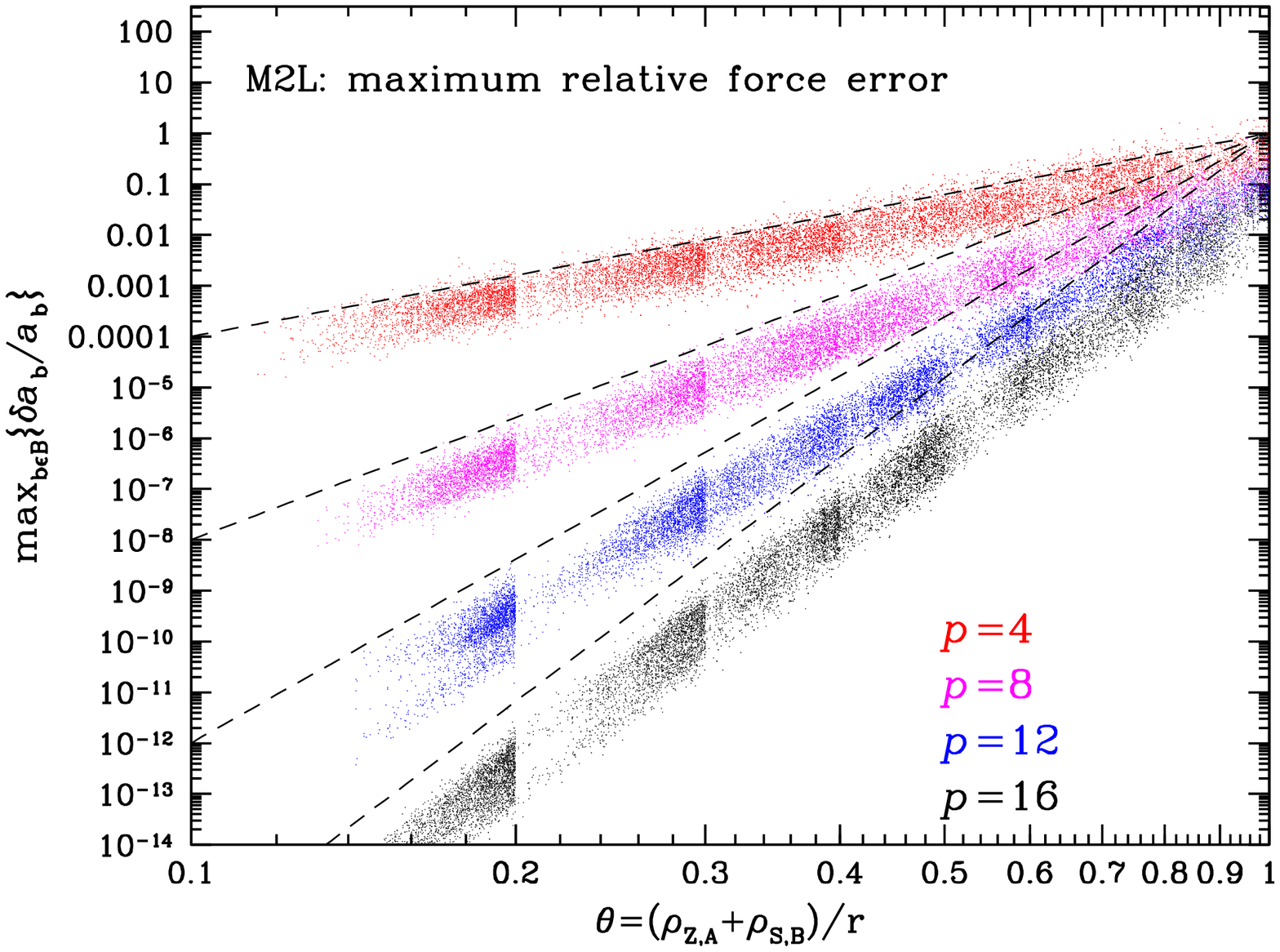}}\hspace{15mm}
    \resizebox{68mm}{!}{\includegraphics{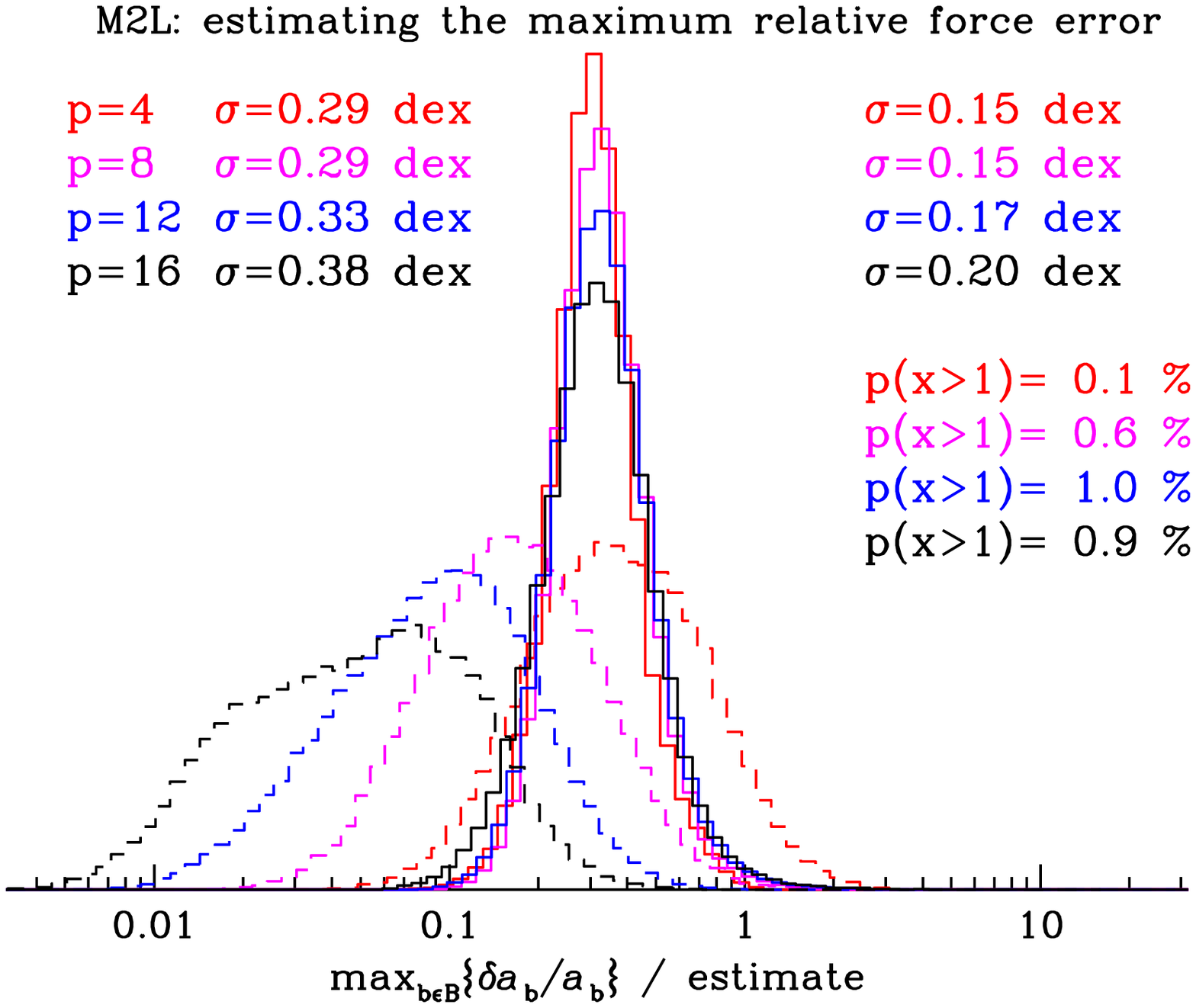}}
  \end{center}
  \begin{minipage}[t]{\columnwidth}
    \caption{Results of the experiments of Section~\ref{sec:errors:m2l}.  Each
      point represents a single cell$\,\to\,$cell interaction with expansion
      order $p$ as indicated. The maximum of the relative (bottom) and absolute
      force error (top; normalised to the mean force $M_{A}/r^2$) within the
      sink cell is plotted against the opening angle $\theta$. The curves in the
      top panel correspond to the relation~(\ref{eq:err:bound}), while the
      dashed lines are the power laws $\theta^p$.
      \label{fig:M2L:err}
    }
  \end{minipage}
  \hfill
  \begin{minipage}[t]{\columnwidth}
    \caption{Distributions of the ratio of the maximum absolute (top)
      and relative (bottom) force error to either the simple
      estimates~(\ref{eq:err:simple}) (dashed, $\sigma$ on the left) or the
      improved estimates~(\ref{eq:err:approx}) (solid, $\sigma$ on the right)
      for the cell-cell interactions of Fig.~\ref{fig:M2L:err}. The reported
      probabilities for error underestimation refer to the improved estimates.
      Vertical scale is linear in the number of interactions.
      \label{fig:M2L:mac}
    }
 \end{minipage}
\end{figure*}
With the simple criterion~(\ref{eq:mac:theta}) the multipole expansion is
guaranteed to converge and have bounded errors\footnote{The original definition
  used in the tree code of \cite{BarnesHut1986} did not ensure bounded errors,
  causing the infamous `exploding galaxies' bug first reported by
  \cite{SalmonWarren1994}.}. Cell$\,\to\,$cell interactions with $N_AN_B<p^3$,
cell$\,\to\,$particle interactions with $N_C<4p^2$, and particle$\,\to\,$cell
interactions with $N_C<p^2$ are ignored, because direct summation is faster than
FMM and will be preferred in a practical application. For the remaining
well-separated interactions, the accelerations of all particles within the sink
cell and due to all particles within the source cell are computed using both FMM
and direct summation (with 64-bit precision for both). I then evaluate for each
sink particle the acceleration error
\begin{equation} \label{eq:err:def}
  \delta a \equiv |\vec{a}_{\mathrm{fmm}}-\vec{a}_{\mathrm{true}}|
\end{equation}
with $\vec{a}_{\mathrm{true}}$ obtained by direct summation in double
precision.

\subsection{Cell-cell interactions}
\label{sec:errors:m2l}
Cell-cell interactions involve the M2L kernel of the
$\mathrm{P2M}+\mathrm{[M2M]}+\mathrm{M2L}+\mathrm{[L2L]}+\mathrm{L2P}$ chain of
kernels. They are by far the most common and most important of all interactions
encountered in the dual tree walk. For a random subset of cell-cell interactions
generated by my experiments, the top panel of Fig.~\ref{fig:M2L:err} plots the
maximum (over all particles within the sink cell) of $\delta a$ normalised by
the average acceleration $M_{A}/r^2$ against $\theta$, while the bottom panel
plots the maximum relative force error $\delta a/a$. As expected, the errors
decrease with smaller $\theta$ and increasing $p$, though there is substantial
scatter at any given $\theta$ and $p$. At $\theta\sim1$, the expansion order has
little effect on the errors, implying that $\theta\ll1$ is required for small
errors.

\begin{figure*}
  \begin{center}
    \resizebox{78mm}{!}{\includegraphics{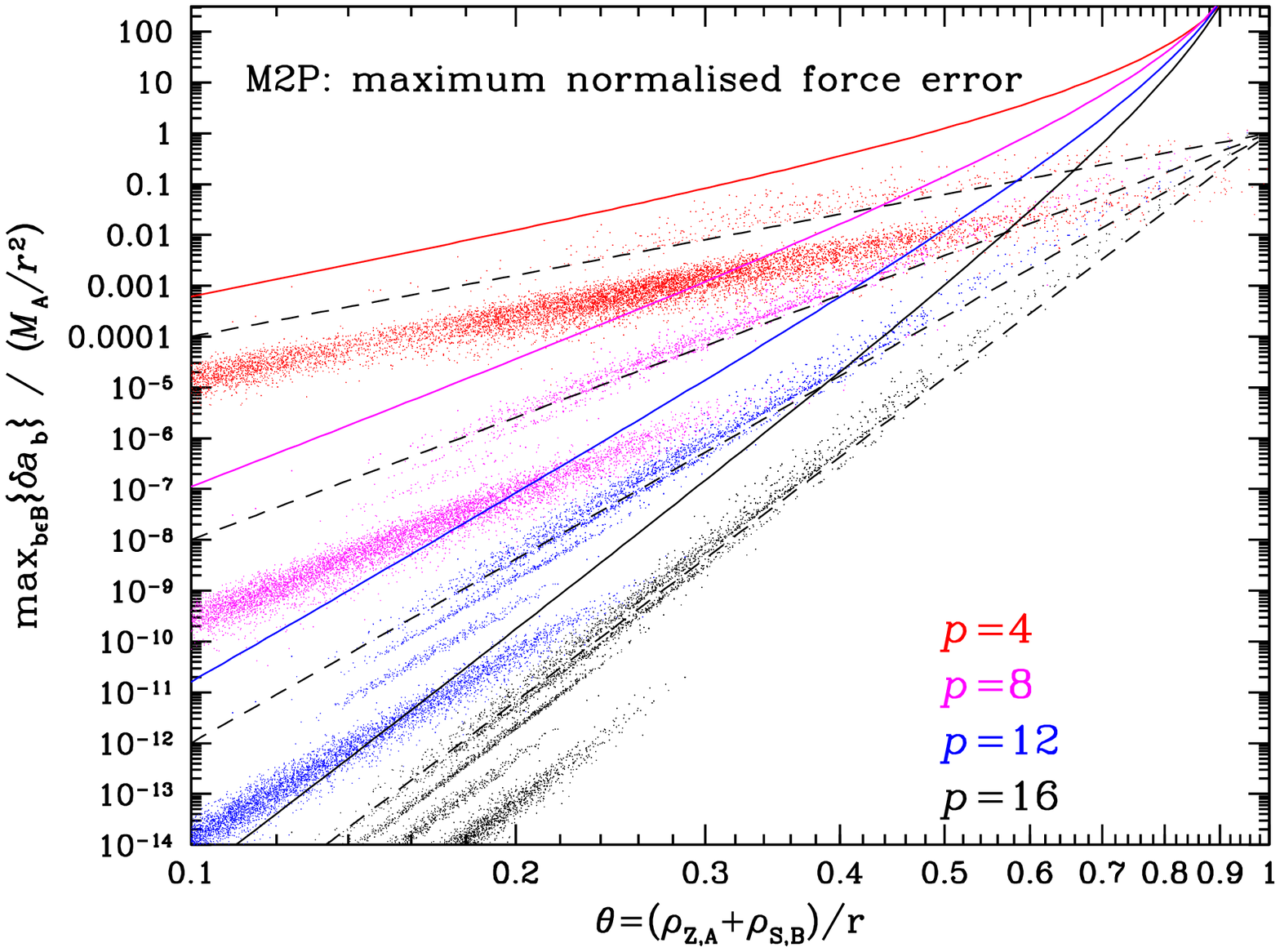}}\hspace{15mm}
    \resizebox{68mm}{!}{\includegraphics{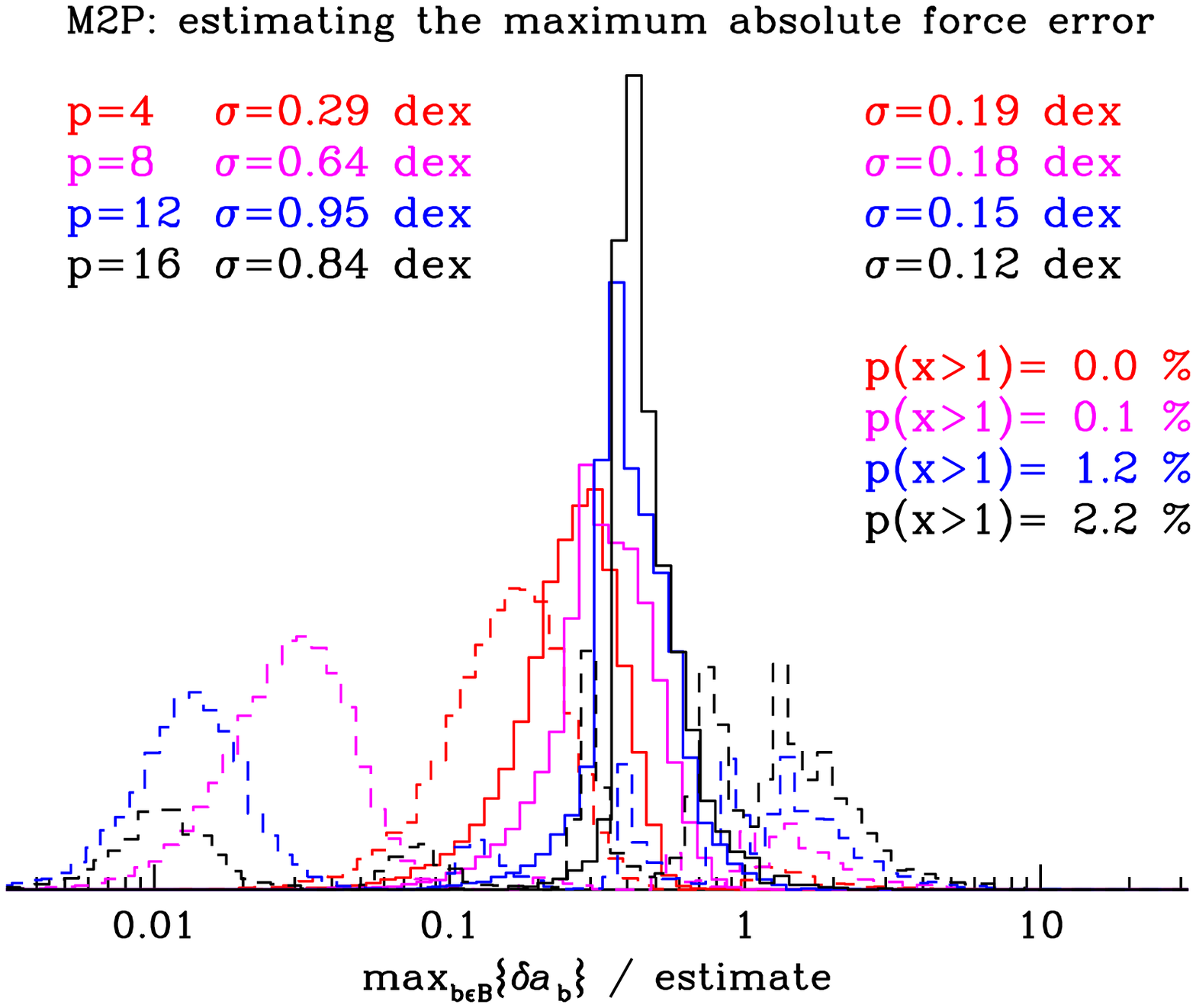}}\\[3mm]
    \resizebox{78mm}{!}{\includegraphics{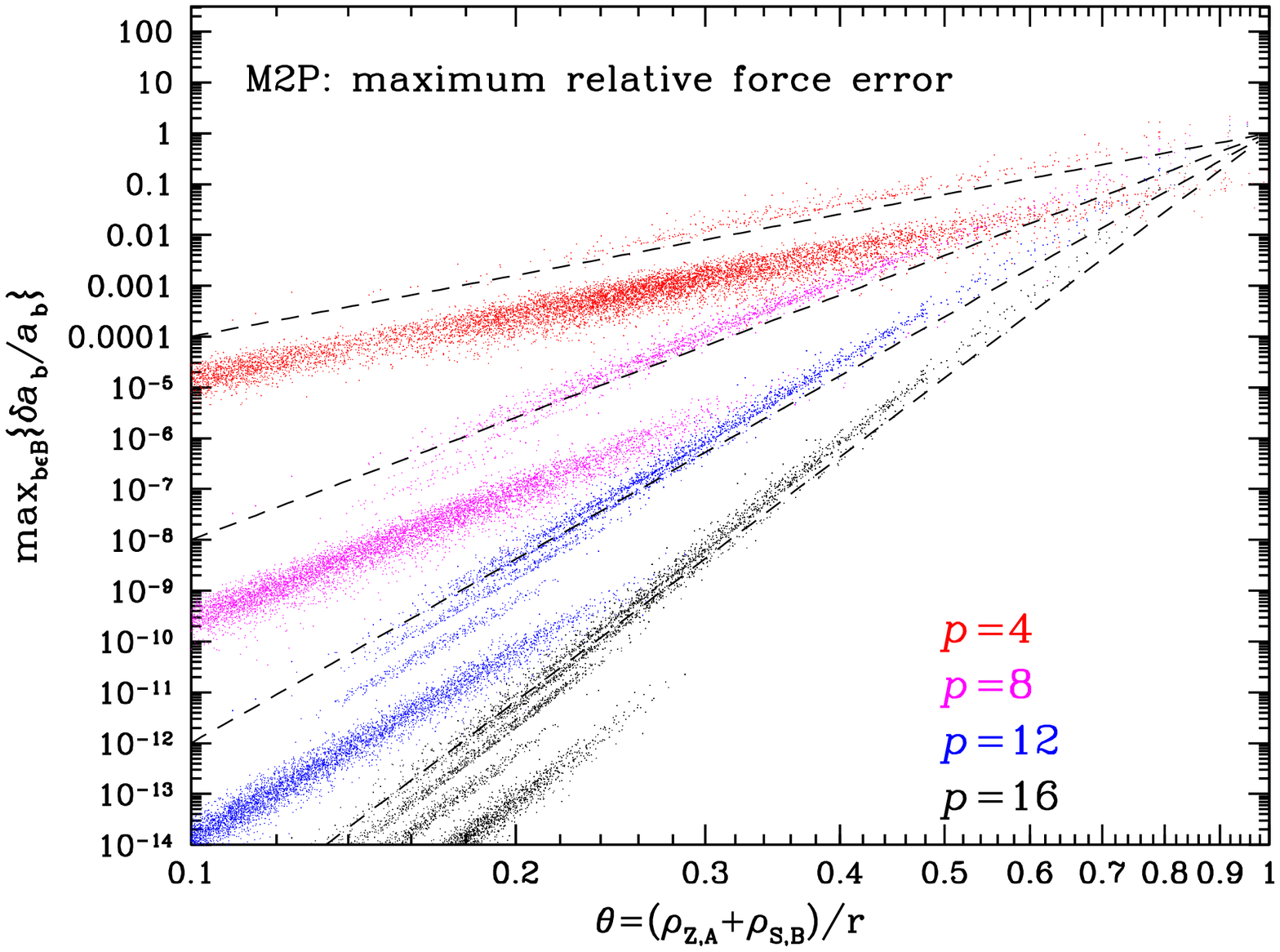}}\hspace{15mm}
    \resizebox{68mm}{!}{\includegraphics{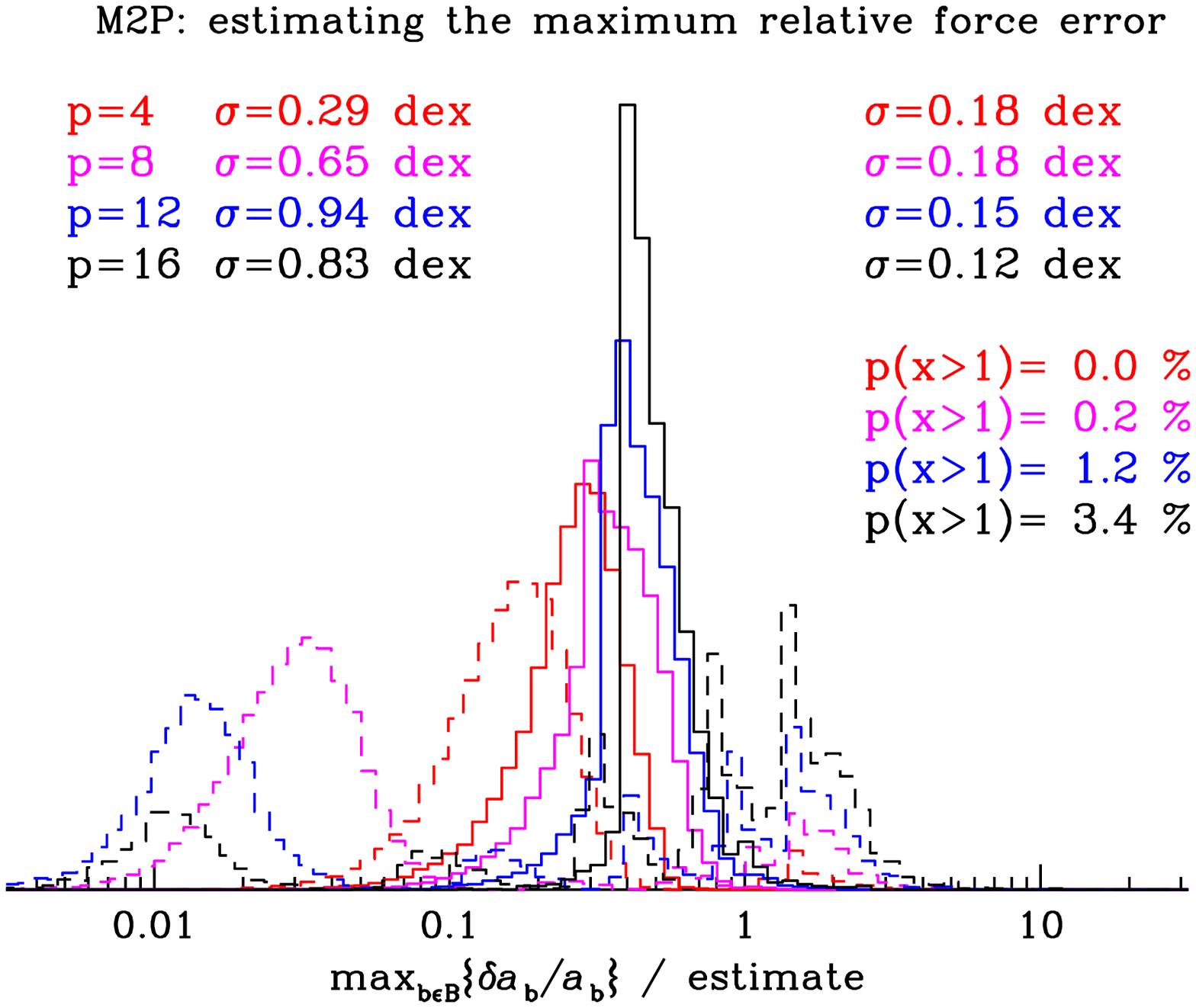}}
  \end{center}
  \caption{As Fig.~\ref{fig:M2L:err} (left) and Fig.~\ref{fig:M2L:mac} (right)
    but for cell$\,\to\,$particle interactions, see
    Section~\ref{sec:errors:m2p}.
    \label{fig:M2P:err}
  }
\end{figure*}
\subsubsection{Comparing with simple error estimates}
The approximation error from a single FMM interaction with $\theta<1$ has the
theoretical strict upper bound \citep{Dehnen2002}
\begin{equation} \label{eq:err:bound}
  \delta a \le
  \frac{M_{A}}{r^2} \frac{(p+1)\theta^p}{(1-\theta)^2},
\end{equation}
which is plotted as thin curves in the top panel of
Fig.~\ref{fig:M2L:err}. Obviously, this upper bound is satisfied, but typically
it is 10-100 times larger than the actual largest error.

\begin{figure*}
  \begin{center}
    \resizebox{78mm}{!}{\includegraphics{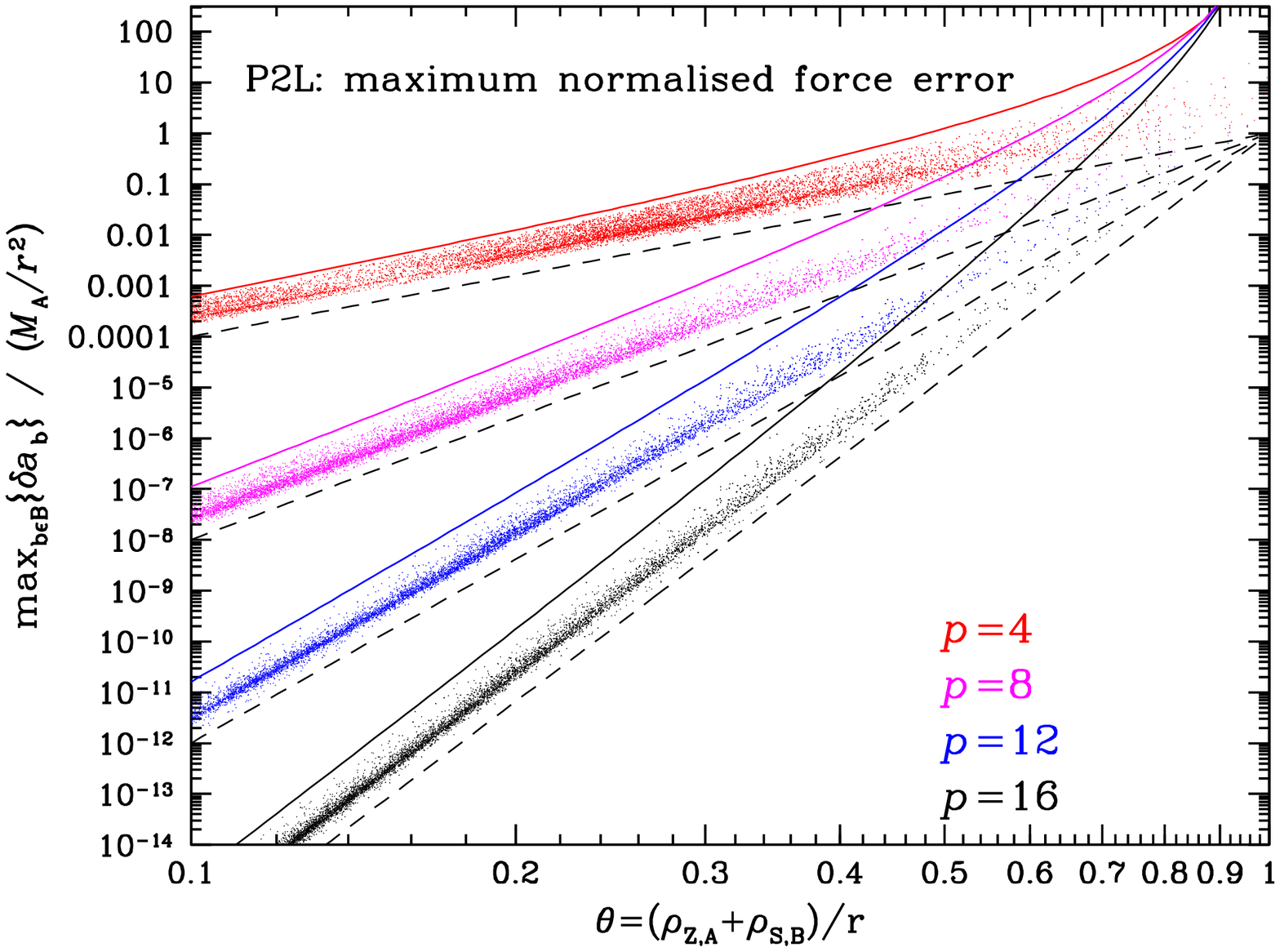}}\hspace{15mm}
    \resizebox{68mm}{!}{\includegraphics{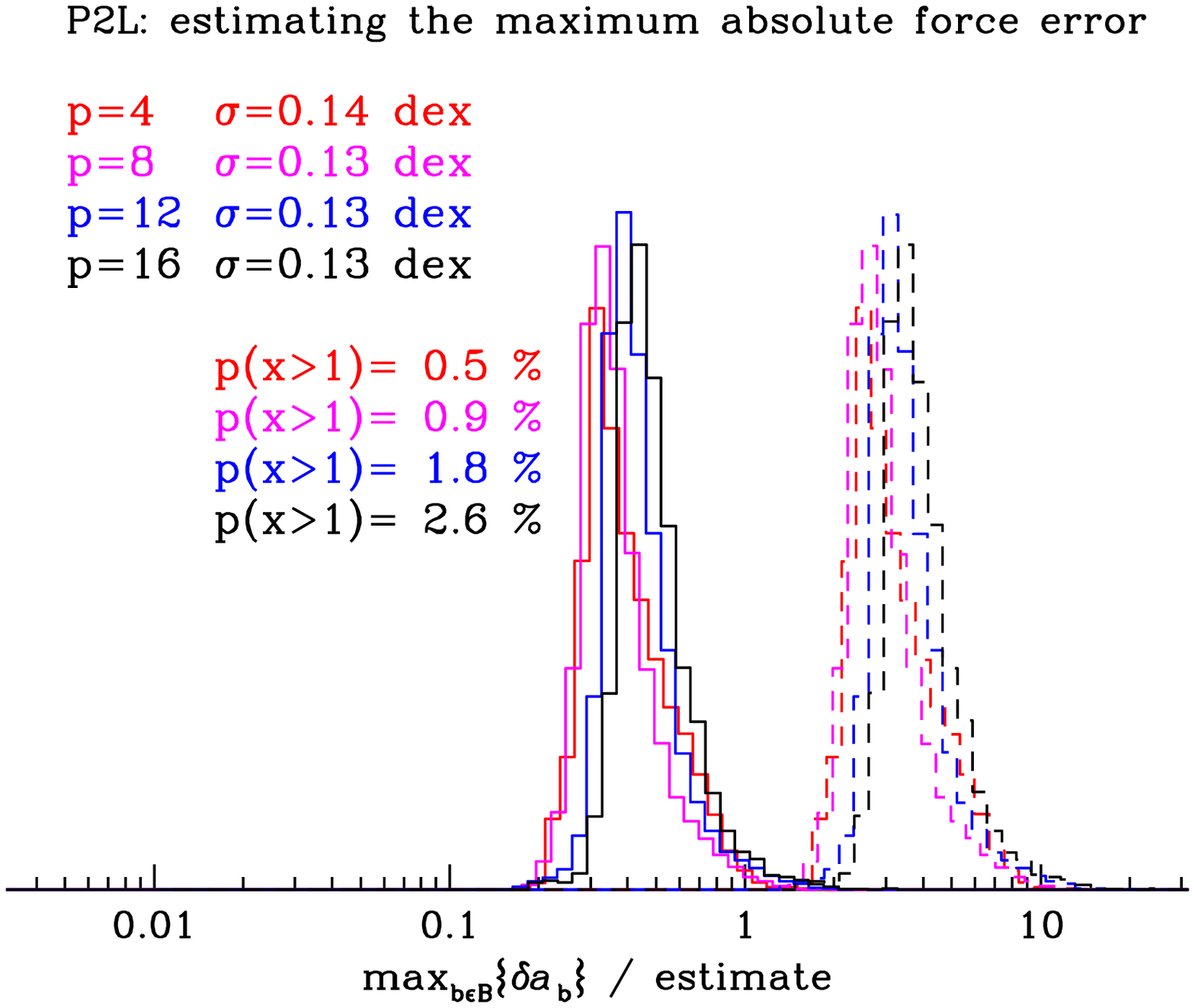}}\\[3mm]
    \resizebox{78mm}{!}{\includegraphics{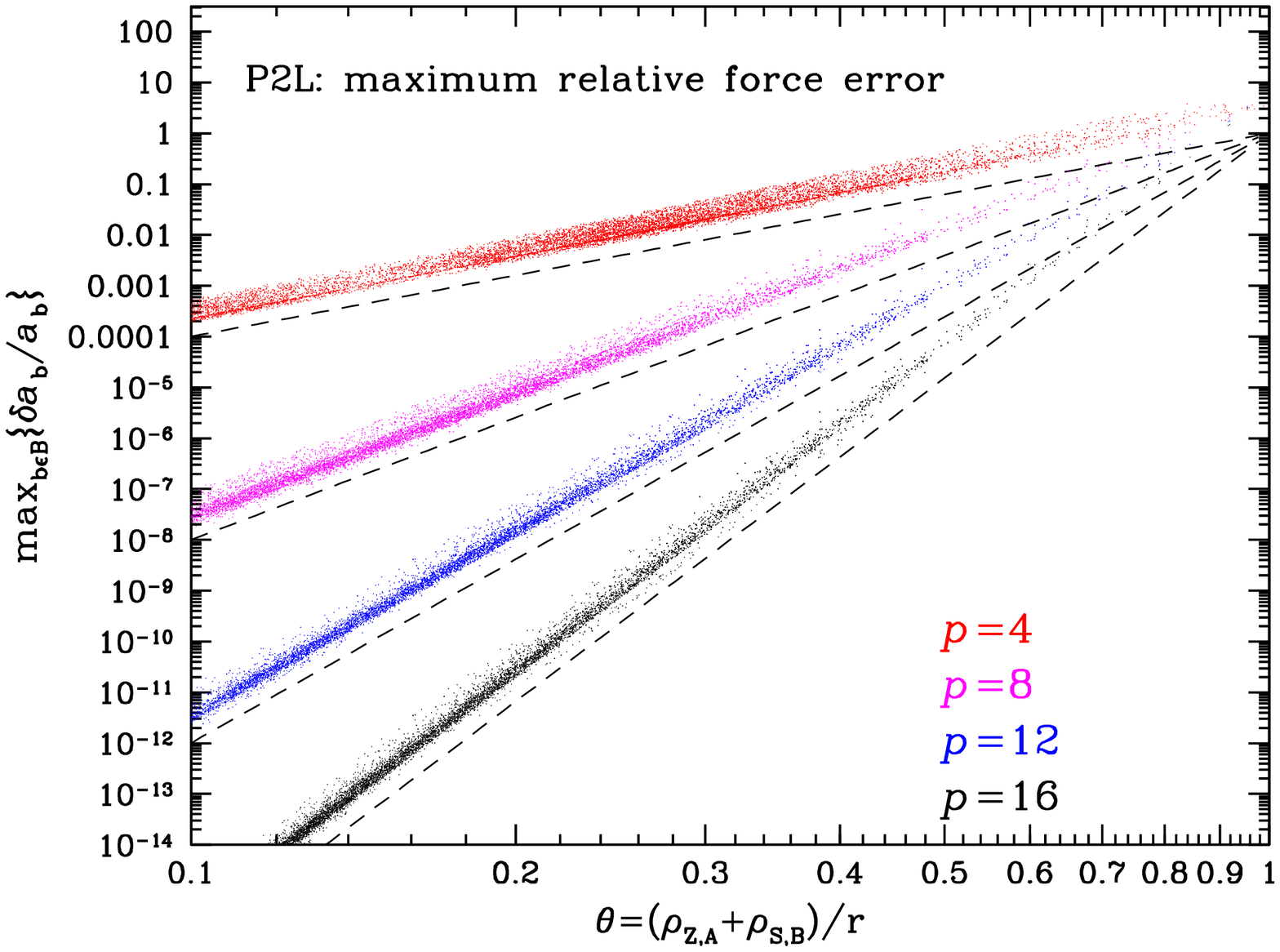}}\hspace{15mm}
    \resizebox{68mm}{!}{\includegraphics{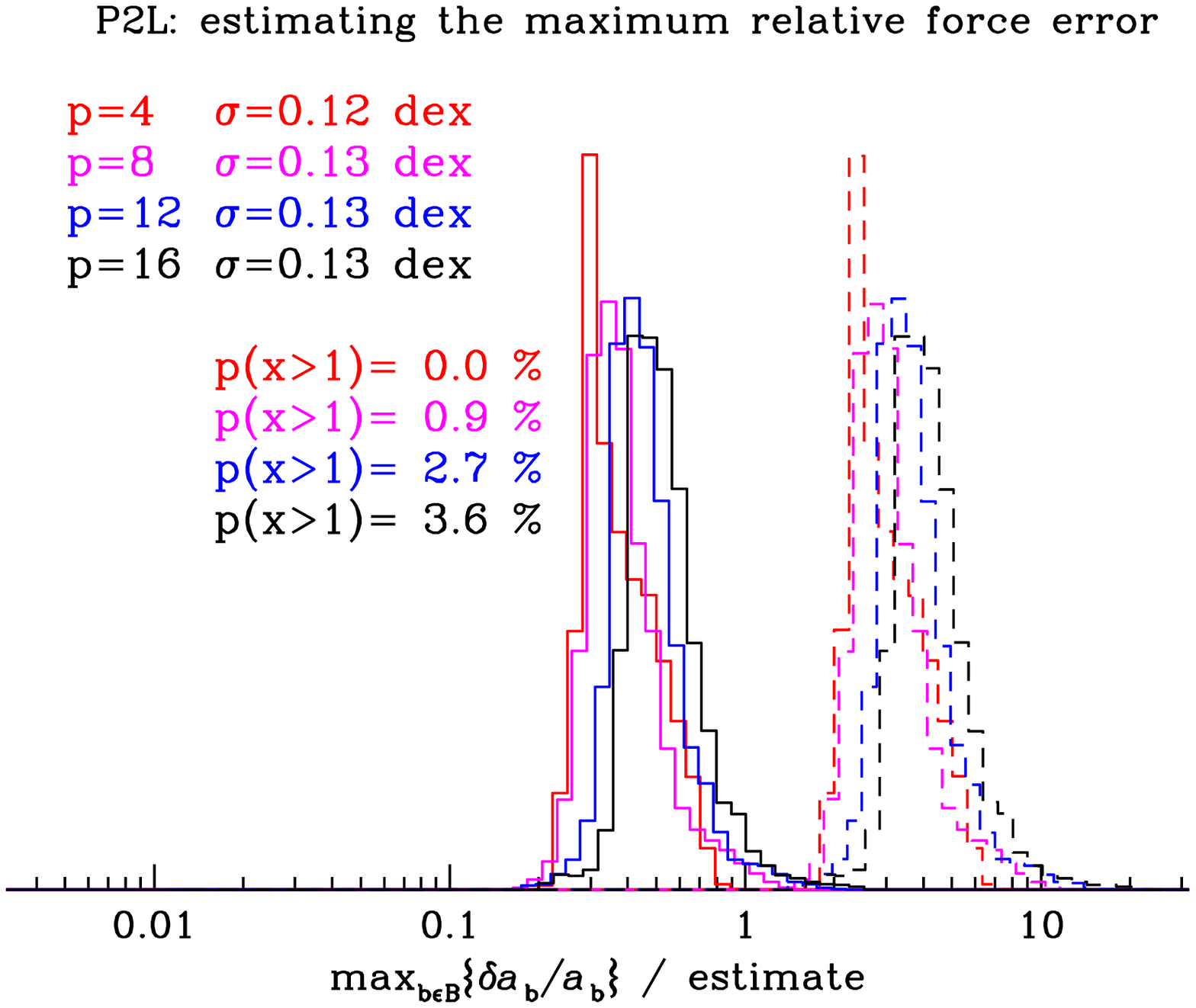}}
  \end{center}
  \caption{As Fig.~\ref{fig:M2L:err} (left) and Fig.~\ref{fig:M2L:mac} (right)
    but for particle$\,\to\,$cell interactions, see
    Section~\ref{sec:errors:p2l}.
    \label{fig:P2L:err}
  }
\end{figure*}

Moreover, equation~(\ref{eq:err:bound}) predicts diverging errors for
$\theta\to1$, while the actual errors behave much nicer. This is presumably
because diverging errors only occur for rare sink positions combined with
extreme source distributions (such as all particles concentrated near one point
at the edge of the source sphere), which are not realised in these experiments
(but see Fig.~\ref{fig:P2L:err}).

Fig.~\ref{fig:M2L:err} also shows as dashed lines the simple power laws
$\theta^p$, which give closer, though not strict, bounds
\begin{equation} \label{eq:err:simple}
  \delta a  \lesssim \theta^p\,M_{A}/r^2
  \quad\text{and}\quad
  \delta a/a \lesssim \theta^p
\end{equation}
to the actual errors. 

\subsubsection{Better error estimates}
The simple error estimates~(\ref{eq:err:simple}) are still quite inaccurate: the
maximal error is often much smaller (see also the dashed histograms in
Fig.~\ref{fig:M2L:mac}). The offsets in Fig.~\ref{fig:M2L:err} of $\theta^p$
from the actual errors increase with $p$. This effect vanishes if the same limit
for $N_AN_B$ is used for all $p$, suggesting that it is caused by smoother
distributions for larger numbers $N_A$ of sources. Indeed, if I simply divide
the estimates~(\ref{eq:err:simple}) by $\sqrt{N_A}$ the scatter of the residuals
is much reduced, but a systematic trend with $p$ remains.

However, there is more information about the distribution of sources than merely
their number: their multipole moments $\mathcal{M}_n^m$ for $n\le p$. In order
to incorporate this information into an error estimate, I first compute for
each cell the multipole power
\begin{equation} \label{eq:def:PA}
  \mathcal{P}_{n}^2 \equiv  
  \sum_{m=-n}^n (n-m)!(n+m)!
  \left|\mathcal{M}_n^m(\vec{z})\right|^2.
\end{equation}
By design these (i) satisfy $\mathcal{P}_{n,A}^{}\le
M_{A}^{}\,\rho_{\vec{z},A}^n$ for any distribution of sources; (ii) are
invariant under rotation (of the coordinate system) and hence independent of the
interaction direction; and (iii) provide an upper bound for the amplitude of the
multipole: $|\mathcal{M}_n^m(\vec{z})|\le\mathcal{P}_{n}/n!$. Having computed
$\mathcal{P}_{n}$ for each source cell, one can evaluate
\begin{equation} \label{eq:def:EAB}
  E_{A\to B} \equiv\frac{1}{M_{A}}
  \sum_{k=0}^{p}  \binom{p}{k}\,
  \frac{\mathcal{P}_{k,A}^{}\,\rho_{\vec{s},B}^{p-k}}{r^{p}}.
\end{equation}
with $\mathcal{O}(p)$
operations.
Note that $E_{A\to B}\le\theta^p$ with equality only for
$\mathcal{P}_{n,A}=M_{A}\,\rho_{A}^n$. The new error estimates are then
\begin{equation}
  \label{eq:err:approx}
  \delta a \lesssim \tilde{E}_{A\to B} \frac{M_{A}}{r^2}
  \quad\text{and}\quad
  \frac{\delta a}{a} \lesssim \tilde{E}_{A\to B}
\end{equation}
with
\begin{equation}
  \label{eq:err:tilde}
  \tilde{E}_{A\to B} = 
  \frac{8\max\{\rho_{\vec{z},A},\rho_{\vec{s},B}\}}
       {\rho_{\vec{z},A}+\rho_{\vec{s},B}}\, E_{A\to B}.
\end{equation}
In Fig.~\ref{fig:M2L:mac}, these new error estimates are compared with the
simple estimates~(\ref{eq:err:simple}) of the last subsection by displaying the
distributions of the ratio of the actual maximum error to these estimates. The
main difference between the two sets of estimators is their accuracy: there is
much less scatter for the new (solid histograms) than for the old estimators
(dashed). Consequently, there are hardly any interactions for which the
force error is overestimated by more than a factor ten, while the simple
estimators~(\ref{eq:err:simple}) overestimated the force error by more than that
for many interactions, in particular at large $p$. Another remarkable property
of the new error estimator is its consistency with respect to expansion order:
there is no systematic drift with expansion order.

The number of underestimated force errors (abscissa $>1$ in
Fig.~\ref{fig:M2L:mac}) is small but there is a clear tail of underestimated
absolute errors (top panel). As this is not present for the relative errors, it
must be caused by the deviation of the acceleration from the mean
$M_{A}/r^2$. Indeed, the maximum error is expected to occur on the side of the
sink towards the source, where the acceleration is larger, about
$M_{A}/(r-\rho_{\vec{s},B})^2$. When accounting for this by simply replacing
$r$ in~(\ref{eq:err:approx}) with $r-\rho_{\vec{s},B}$, the tail of
underestimated force errors is diminished, but the overall distributions widens
and a tail of overestimated errors appears.

\subsection{Particle-cell interactions}
\label{sec:errors:m2p:p2l}
Just occasionally, the dual tree walk algorithm encounters particle-cell
interactions. Most of them will be computed using direct summation, leaving
only the few with populous cells for the FMM approximation.

For $\text{particle}\to\text{cell}$ and $\text{cell}\to\text{particle}$
interactions the FMM approximation uses the P2L and M2P kernels, respectively.
Because these kernels correspond to the M2L kernel in the limits of
$\rho_{\vec{z},A}\to0$ and $\rho_{\vec{s},B}\to0$, respectively, all the
algebra developed in the previous sub-section still applies.

\subsubsection{Cell $\to$ particle interactions}
\label{sec:errors:m2p}
The left panels of Fig.~\ref{fig:M2P:err} are equivalent to
Fig.~\ref{fig:M2L:err} for cell$\,\to\,$particle interactions (which dominate in
the tree code). The most notable difference to Fig.~\ref{fig:M2L:err} is the
streaky nature of the relations, implying a multi-modal distribution of errors
at any given $\theta$ and $p$, as also evident from the dashed histograms in the
right panels. The cause for this is simply that in an oct-tree cell size is
quantised. In fact, the improved error estimates~(\ref{eq:err:approx}) account
for this effect resulting in narrow mono-modal distributions of error offsets.

\subsubsection{Particle $\to$ cell interactions}
\label{sec:errors:p2l}
The left panels of Fig.~\ref{fig:P2L:err} are equivalent to
Fig.~\ref{fig:M2L:err} for particle$\,\to\,$cell interactions. Clearly, at any
given $\theta$ and $p$, the errors are larger than for any other type of
interactions and are in fact approaching the theoretical limit (solid curves in
the top left panel). What is more, not much can be done about this in terms of
error estimates: since the source is just a particle without inner structure,
the improved estimates~(\ref{eq:err:approx}) are simply a rescaling by a factor
8 from the simple power laws (a simple shift between the dashed and solid
histograms in the right panels). They are nonetheless equally accurate as for
the cell$\,\to\,$cell interactions and suffer from a similar level of force
underestimation (for a few percent of interactions and by less than a factor
two).

\section{Optimising the multipole-acceptance criterion}
\label{sec:mac}
With the improved error estimates in hand, the practical implementation of FMM
for high accuracy can finally be considered. The main questions arising in this
context are:
\goodbreak
\begin{itemize}
  \item what to pick for the expansion centres $\vec{z}$ and $\vec{s}$?
  \item when to consider two cells well-separated?
  \item what expansion order $p$ to use?
\end{itemize}
The possible answers to these questions affect both the computational cost and
the approximation accuracy. Hence, for a given accuracy target, there exists an
optimal choice for all these parameters, in the sense of minimal CPU time (and
memory) consumption. This section deals with the algorithmic aspects of this
problem, i.e.\ the choice for $\vec{z}$ and $\vec{s}$ and the functional form of
the multipole-acceptance criterion. The tuning of the parameters (of the
multipole-acceptance criterion as well as the expansion order) with the aim of
minimal computational effort for a given accuracy is the subject of the next
section.

Astonishingly, this issue of optimal choice for $\vec{z}$ and $\vec{s}$ and the
multipole-acceptance criterion has not been much investigated. Instead,
implementations of multipole methods often employ either of two simple
strategies. The tree-code generally uses a fixed order $p$ and an expansion
centred on the cells' centres of mass, while two cells are considered
well-separated if the simple geometric multipole-acceptance
criterion~(\ref{eq:mac:theta}) is satisfied, such that $\theta_{\mathrm{crit}}$
controls the accuracy.

With traditional FMM, on the other hand, the expansion centres $\vec{z}$ and
$\vec{s}$ are both taken to be the geometric cell centres and two cells are
deemed well-separated as soon as the expansion converges, corresponding to
$\theta_{\mathrm{crit}}=1$. When using hierarchical cubic grids (instead of an
adaptive tree), this is implemented by interacting only between non-neighbouring
cells on the same grid level whose parent cells are neighbours
\citep[e.g.][]{ChengGreengardRokhlin1999}. The accuracy is then only controlled
by the expansion order $p$.

\subsection{Choice of expansion centres $\vec{z}$ and $\vec{s}$}
\label{sec:mac:zc}
As far as I am aware, all existing FMM implementations use the same position for
the multipole and potential expansion centres, i.e.\ $\vec{z}=\vec{s}$, for each
cell. For traditional FMM, these are equal to the geometric cell centres. This
has the benefit of a finite number of possible interaction directions
$\hat{\vec{r}}$, in particular when $\theta_{\mathrm{crit}}=1$, for which the
coefficients $\Theta_n^m(\hat{\vec{r}})$ could be pre-computed. However, the
computation of these coefficients on the fly is often faster than a table
look-up. Moreover, in view of Fig.~\ref{fig:M2L:err} $\theta_{\mathrm{crit}}=1$
appears ill-suited for high accuracy.

In fact, the restriction $\vec{z}=\vec{s}$ reduces the freedom and hence the
potential for optimising the method. Nonetheless, when aiming for low accuracy,
choosing $\vec{z}=\vec{s}=\vec{z}_{\mathrm{com}}$, the cells' centres of mass,
has some advantages. First, the dipoles vanish and the low-order multipoles tend
to be near-minimal. Second, \emph{if} using a mutual version of the algorithm
(when the interactions $A{\,\to\,}B$ and $B{\,\to\,}A$ are done simultaneously),
the computational costs are reduced and the approximated forces satisfy Newton's
third law exactly, i.e.\ $\vec{F}_{\!ab}+\vec{F}_{\!ba}=0$ \citep{Dehnen2002}.

However, in practice there is no benefit from such an exact obedience of
Newton's law, as the total momentum is not exactly conserved, because of
integration errors arising from the fact that the particles have individual time
steps. Moreover, the degree of deviation from exact momentum conservation in
such a case does not reflect the true accumulated force errors. In a more
general method, the approximated forces will deviate from the ideal
$\vec{F}_{\!ab}+\vec{F}_{\!ba}=0$ by an amount comparable to their actual force
errors and the non-conservation of total momentum is somewhat indicative of the
accumulated effect of the force errors (see also
Appendix~\ref{app:error:approx}).

\subsubsection{Choice of the potential expansion centre $\vec{s}$}
\label{sec:mac:z}
The results of Section~\ref{sec:errors}, in particular the functional form of
$E_{A\to B}$ in equation~(\ref{eq:def:EAB}), suggest to choose the potential
expansion centres $\vec{s}$ such that the resulting sink radii $\rho_{\vec{s}}$,
and hence the estimated interaction errors, are minimal. Thus,
$\vec{s}=\vec{z}_{\mathrm{ses}}$, the centre of the \emph{smallest enclosing
  sphere}. Finding the smallest enclosing sphere for a set of $n$ points has
complexity $\mathcal{O}(n)$. Doing this for every sink cell would incur a total
cost of $\mathcal{O}(N\ln N)$ and be prohibitively expensive.

Instead, I use an accurate approximation by finding for each cell the smallest
sphere enclosing the spheres of its grand-daughter cells. This incurs a total
cost of $\mathcal{O}(N)$ and is implemented via the Computational Geometry
Algorithms Library \citep[www.cgal.org,][]{cgal:fghhs-bv-12b}, using an
algorithm of \cite*{MatousekSharirWelzl1996}.

\subsubsection{Choice of the multipole expansion centre $\vec{z}$}
\label{sec:mac:c}
As already mentioned above, setting $\vec{z}=\vec{z}_{\mathrm{com}}$ has some
virtue for low expansion orders $p$. However, for high expansion orders, the
high-order multipoles become ever more important, suggesting that
$\vec{z}=\vec{z}_{\mathrm{ses}}$ may be a better choice. In order to assess the
relative merits of these methods, I repeated the experiments of
Section~\ref{sec:errors} for both methods and compared the resulting maximum
absolute and relative force errors incurred for the \emph{same}
$\text{cell}\to\text{cell}$ interactions (for which the two methods give
different $\theta$).

I found that the errors for the two methods are very similar with an rms
deviation of $\sim0.15\,$dex, but a very small mean deviation. At $p\lesssim8$
there is a trend of more accurate forces for $\vec{z}=\vec{z}_{\mathrm{com}}$,
while at $p\gtrsim8$ $\vec{z}=\vec{z}_{\mathrm{ses}}$ obtains smaller
errors. This trend is simply a consequence of $\mathcal{P}_{k}$ being smaller
for $\vec{z}=\vec{z}_{\mathrm{com}}$ than for $\vec{z}=\vec{z}_{\mathrm{ses}}$
at low $k$ and larger at high $k$. This together with the improved error
estimates~(\ref{eq:err:approx}) also explains that (for an interaction $A\to B$)
$\vec{z}=\vec{z}_{\mathrm{com}}$ tends to give more accurate forces if
$\rho_{\vec{z},A}<\rho_{\vec{s},B}$, while $\vec{z}=\vec{z}_{\mathrm{ses}}$
tends to be more accurate if $\rho_{\vec{z},A}>\rho_{\vec{s},B}$.

\begin{figure}
  \begin{center}
    \resizebox{\columnwidth}{!}{\includegraphics{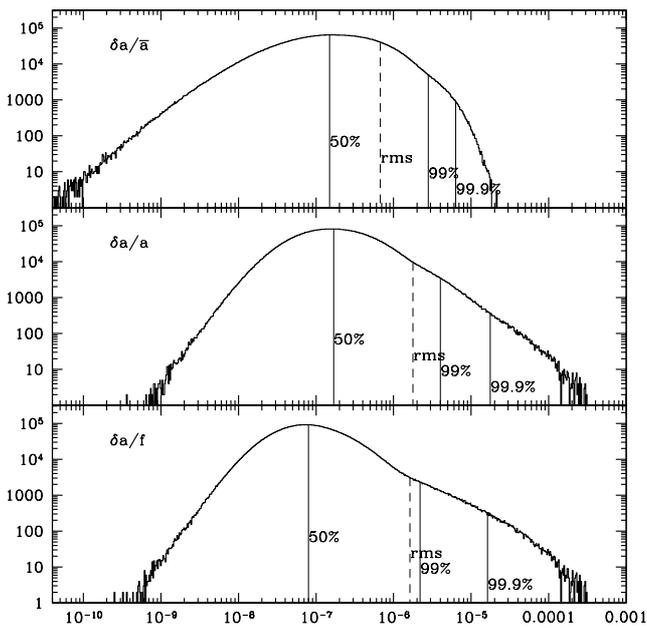}}
  \end{center}
  \caption{Similar to Fig.~\ref{fig:da:gpu}, but for $N=10^7$ particles and
    accelerations obtained by FMM using expansion order $p=8$ and
    multipole-acceptance criterion $\rho_{\vec{z},A} + \rho_{\vec{s},B} < r
    \theta_{\mathrm{crit}}$ (equation~\ref{eq:mac:theta}) with
    $\theta_{\mathrm{crit}} = 0.4$. Bins are 0.01\,dex wide.
    \label{fig:da:simple}
  }
\end{figure}
\subsection{A simple FMM implementation} \label{sec:err:simple}
Let us first experiment with an implementation that uses the simple
multipole-acceptance criterion~(\ref{eq:mac:theta}) and a fixed expansion order
$p$. This is the standard choice for the tree code and as such implemented in
many gravity solvers used in astrophysics. The computational costs of such an
implementation roughly scale as $p^\alpha/\theta_{\mathrm{crit}}^3$ with
$\alpha\sim2.3$, because the number of interactions increases as
$\theta_{\mathrm{crit}}^{-3}$ for large $N$, while the cost for one is $\propto
p^{2.3}$. Together with the simple error estimate~(\ref{eq:err:simple}), this
means that \emph{if} one aims each FMM interaction to satisfy $\delta
a/a<\epsilon$, then the minimum cost for fixed $\epsilon$ occurs for
\[
  \theta_{\mathrm{crit}} = \mathrm{e}^{-\alpha/3} \approx 0.46
\]
Thus, the optimal opening angle is independent of $p$. The accuracy is then
controlled by the expansion order, requiring $p\gtrsim16$ for $\epsilon<10^{-8}$
(according to Fig.~\ref{fig:M2L:err}). The computational costs rise roughly like
$|\ln\epsilon|^\alpha$ with decreasing $\epsilon$.

I applied the FMM method with $\vec{z}=\vec{z}_{\mathrm{ses}}$, expansion order
$p=8$, and $\theta_{\mathrm{crit}}=0.4$ to $N=10^7$ equal-mass particles drawn
from a Plummer sphere. Fig.~\ref{fig:da:simple} plots the resulting
distributions of absolute (top), relative (middle), and scaled (bottom)
acceleration errors. All three distributions are mono-modal, but very wide, much
wider than those obtain from GPU-based direct summation
(Fig.~\ref{fig:da:gpu}). In particular, there are extended tails towards very
large relative or scaled errors, containing only $\lesssim1\%$ of the particles
but reaching up to 1000 times the median error. These tails are due to particles
at large radii and, for the relative errors only, also at small radii (see the
discussion in Section~\ref{sec:error:scaled}).

There are two main effects responsible for these properties of the error
distributions. First, errors from a single FMM interaction follow a distribution
with variance of $1-2$ dex. The maximum errors reported in
Section~\ref{sec:errors} only occur for particles near the edges and corners of
the sink cell, while most have smaller errors. Moreover, the force errors due to
FMM interactions of the same sink cell with source cells in opposing directions
tend to partially cancel rather than add up. Both explain why the median errors
reported in Fig.~\ref{fig:da:simple} are much smaller than the maximum relative
error incurred by a single cell$\,\to\,$cell interaction, which according to
Fig.~\ref{fig:M2L:err} is $\sim10^{-4}$.

\begin{figure*}
  \begin{center}
    \resizebox{\columnwidth}{!}{\includegraphics{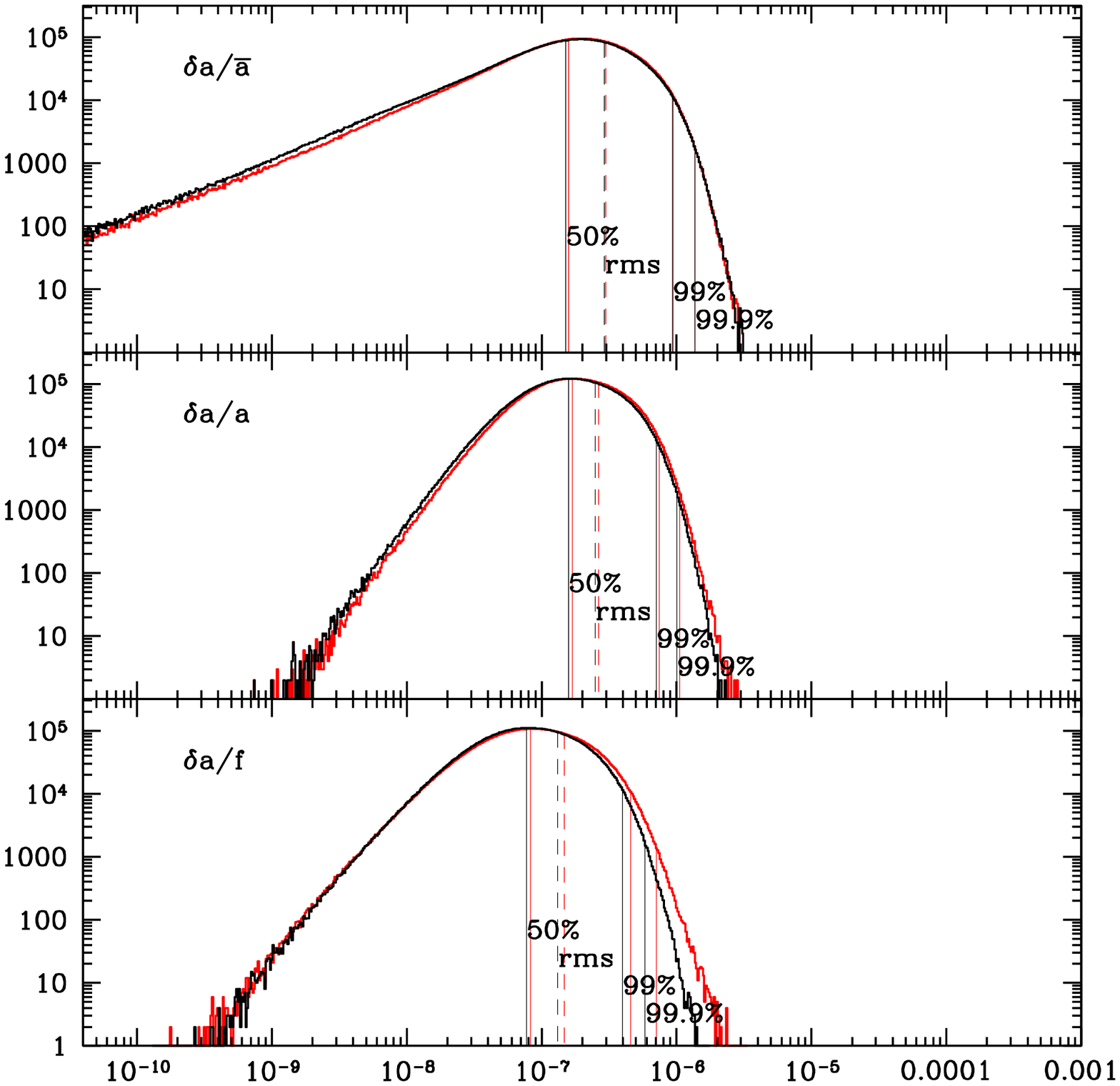}}\hfill
    \resizebox{\columnwidth}{!}{\includegraphics{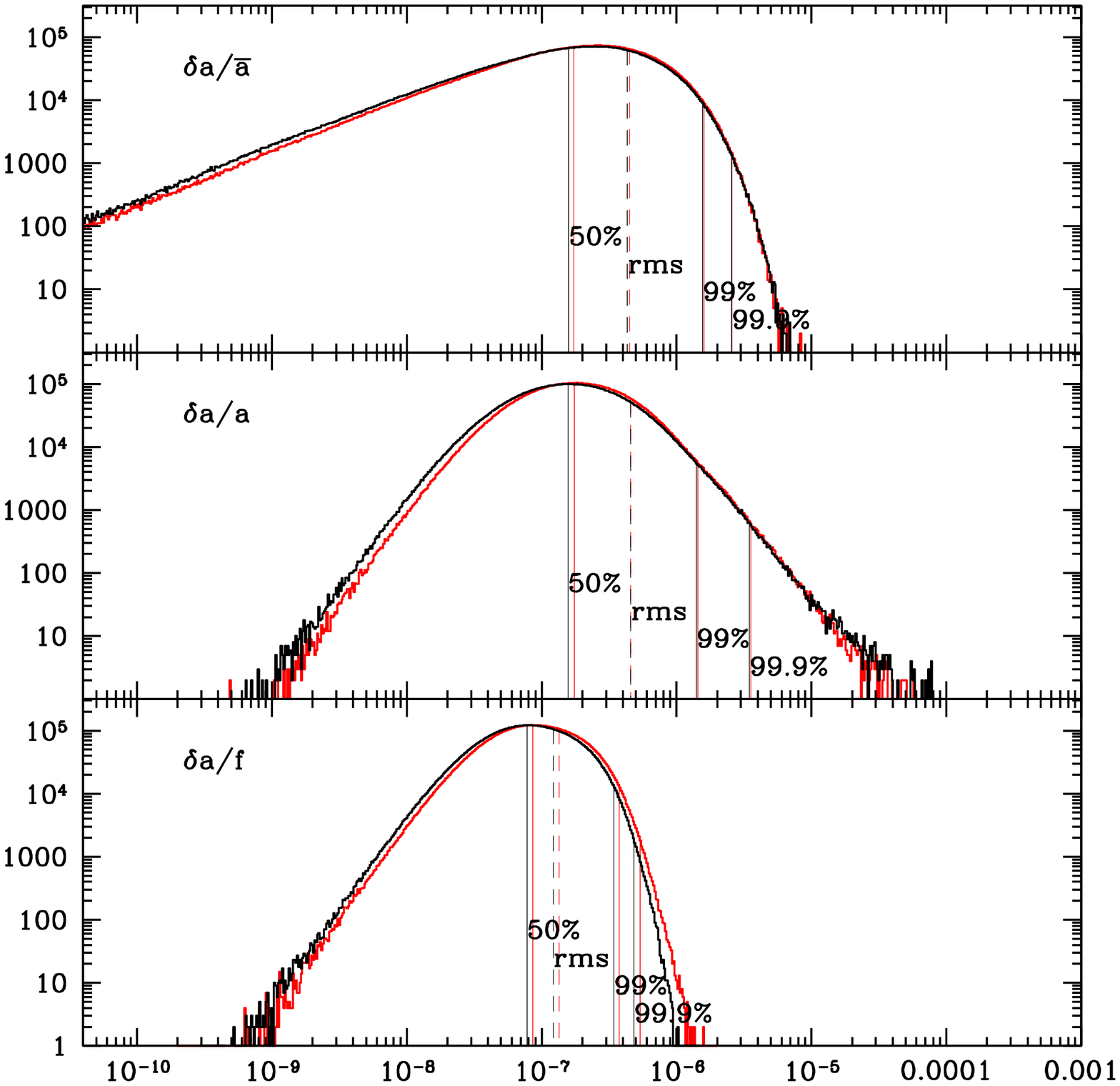}}
  \end{center}
  \caption{Same as Fig.~\ref{fig:da:simple} but for the multipole-acceptance
    criterion~(\ref{eq:mac:better:a}) with $\epsilon=2\times10^{-7}$ (left) or
    (\ref{eq:mac:better:f}) with $\epsilon=10^{-7}$ (right). The values for
    $a$ and $f$ are either taken from the direct-summation run (black), or
    obtained by low-oder FMM (red, see section~\ref{sec:err:improve:fmm}). In
    all cases the computational effort is similar to that of the FMM run shown
    in Fig.~\ref{fig:da:simple} (since $a\le f$
    criterion~(\ref{eq:mac:better:a}) gives a tighter constraint than
    (\ref{eq:mac:better:f}) and hence requires larger $\epsilon$ for the same
    computational effort).
    \label{fig:da:better}
  }
\end{figure*}
More important is a second effect: the final force errors are not the sum of the
relative errors of individual FMM interactions, which are controlled by the
simple multipole-acceptance criterion, but of their \emph{absolute} errors
$\delta a$. Since, according to equation~(\ref{eq:err:simple}), $\delta a \sim
\theta^pM_A/r^2 \sim \theta^{p+2}M_A/4\rho_{\vec{z},A}^2$, the FMM interactions
with cells of large surface density $M/\rho_{\vec{z}}^2$ dominate the error
budget. In fact, the particles at very large radii have $\delta a/a\approx\delta
a/f\sim10^{-4}$, exactly as expected from a few FMM interactions with near
maximal errors.

\subsection{Towards better multipole-acceptance criteria} \label{sec:better}
This discussion suggests that multipole-acceptance criteria which balance the
absolute force errors of individual FMM interactions are preferable. When
working with the simple estimators~(\ref{eq:err:simple}) or the error
bound~(\ref{eq:err:bound}), this leads to critical opening angles which depend
on the properties of the interacting cells, such as their mass or surface
density. 

Such an approach can indeed be made to work \citep{Dehnen2002}, but the aim here
is to go beyond that and use the improved error estimates~(\ref{eq:err:approx}).
This results in the multipole-acceptance criteria
\begin{subequations} \label{eqs:mac:better}
  \begin{eqnarray}
    \label{eq:mac:better:a}
    \theta < 1 \,\, &\wedge& \,\, \tilde{E}_{A\to B} \frac{M_A}{r^2} < \epsilon
    \,\min_{b\in B}\{a_b\}, \\
    \label{eq:mac:better:f}
    \theta < 1 \,\, &\wedge& \,\, \tilde{E}_{A\to B} \frac{M_A}{r^2} < \epsilon
    \,\min_{b\in B}\{f_b\}
  \end{eqnarray}
\end{subequations}
with the aim to obtain $\delta a/a \lesssim \epsilon$ and $\delta a/f \lesssim
\epsilon$, respectively.

The black histograms in Fig.~\ref{fig:da:better} show the error distributions
resulting from these criteria, when the values for $a_b$ and $f_b$ used in
equations~(\ref{eqs:mac:better}) have been taken from the direct-summation
comparison run. The distributions for $\delta a/a$ in the left and $\delta a/f$
in both panels are remarkably narrow with a median error $\sim\epsilon$ as
targeted, a steep truncation towards large errors, and a maximum error
$\sim10\epsilon$. The tail of large $\delta a/a$ in the right panel is due to
particles at small radii, for which $a\ll f$ such that
criterion~(\ref{eq:mac:better:f}) allows large $\delta a/a$.

The difference between these error distributions and those shown in
Fig.~\ref{fig:da:simple} and resulting from the simple geometric
multipole-acceptance criterion~(\ref{eq:mac:theta}) is remarkable. While the
median errors are comparable, the criteria~(\ref{eqs:mac:better}) do not
produce extended tails of large errors of the quantity controlled ($\delta
a/a$ in left and $\delta a/f$ in the right panels of
Fig.~\ref{fig:da:better}), and the maximum errors are more than 2 orders of
magnitude smaller. What is more, the tails towards small errors have also been
somewhat reduced, indicating that the improved criterion avoids overly
accurate individual FMM interactions.

This improvement has been achieved without increasing the overall
computational effort, but by carefully considering the error contribution from
each approximated interaction.

\subsection{Practical multipole-acceptance criteria} \label{sec:practics}
In a real application one has, of course, no a priori knowledge of $a_b$ or
$f_b$ for any particle and must instead use something else in the
multipole-acceptance criteria~(\ref{eqs:mac:better}). In some situations, a
suitable scale can be gleaned from the properties of the system modelled. For
example, if simulating a star cluster of known mass profile $M(r)$ and centre
$\vec{x}_0$, one may simply use $a_b\sim GM(r_b)r_b^{-2}$ with
$r_b=|\vec{x}_b-\vec{x}_0|$. I now consider other options.

\subsubsection{Using accelerations from the previous time step}
\label{sec:err:improve:acc}
Employing the accelerations $\vec{a}_b$ from the previous time step in
equation~(\ref{eq:mac:better:a}) requires no extra computations. However, it
means that the gravity solver is not self-contained, but requires some starter
to get the initial accelerations.

Also, using information from the previous time step subtly introduces an
artificial arrow of time into the simulation, because $\delta
a_{\mathrm{new}}<\epsilon a_{\mathrm{old}}$ implies $\delta
a_{\mathrm{new}}/a_{\mathrm{new}} <\epsilon\,
a_{\mathrm{old}}/a_{\mathrm{new}}$. Hence, a particle moving in a direction of
increasing acceleration has, on average, smaller $\delta a/a$ than when moving
in the opposite direction, or in reversed time. However, the time integration
methods currently employed almost exclusively in $N$-body simulations of
collisional stellar dynamics are irreversible and introduce their own arrow of
time. This suggests, that the additional breach of time symmetry by the
magnitude (not the direction) of the force error may not be a serious problem in
practice\footnote{The situation is different for $N$-body simulations of
  collisionless stellar dynamics, where reversible integrators are used and the
  accepted force errors, and thus their time asymmetries, are significantly
  larger.}.

\subsubsection{Estimating $a_b$ or $f_b$ using low-order FMM}
\label{sec:err:improve:fmm}
As Section~\ref{sec:errors} has shown, the error estimate $\tilde{E}_{A\to B}$
used in the multipole-acceptance criteria~(\ref{eqs:mac:better}) still has
significant uncertainty, and using highly accurate values for $a_b$ or $f_b$
in equations~(\ref{eqs:mac:better}) is unnecessary. Instead, rough estimates
should suffice. Such estimates can be obtained via a low-order FMM.  This
amounts to running the FMM twice: once with a simple multipole-acceptance
criterion to obtain rough estimates for $a_b$ or $f_b$, and then again using
the sophisticated criteria~(\ref{eqs:mac:better}) employing the results of the
first run.

The acceleration scale $f$ (defined in equation~\ref{eq:f}) is similar to the
gravitational potential~(\ref{eq:P2P}), except that its Greens function is
$|\vec{r}|^{-2}$. This implies that it too can be estimated using FMM, albeit
not using an explicitly harmonic formulation.

I implemented both options, estimating $\vec{a}$ or $f$ via FMM, using the
lowest possible order ($p=0$ for $f$ and $p=1$ for gravity---recall that
$\vec{a}=\vec{\nabla}\Psi$ is approximated at one order lower than the potential
$\Psi$) and multipole-acceptance criterion $\theta<1$. To this end, I use
$\vec{s}=\vec{z}=\vec{z}_{\mathrm{com}}$ and a mutual version of the dual tree
walk. The resulting estimates for $f$ or $a=|\vec{a}|$ have rms relative errors
of $\sim15\%$. The additional computational effort is still much smaller than
that of the high-accuracy approximation of gravity itself, though estimating $f$
is faster because it is a scalar rather than a vector and because no square-root
needs to be calculated.

The distributions of acceleration errors resulting from using these estimates in
equations~(\ref{eqs:mac:better}) are shown in red in Fig.~\ref{fig:da:better}.
They are only very slightly worse than those in black, which have been obtained
using the exact values of $a_b$ and $f_b$ in equations~(\ref{eqs:mac:better}).

\section{Optimising adaptive FMM}
\label{sec:opt}
The previous section provided answers to the first two questions asked at its
beginning, but not to the one after the optimal expansion order $p$. To answer
this question I now report on some experiments, which also provide the actual
computational costs for a given required force accuracy.

All experiments are run on a single compute node with 16 Intel Xeon E5-2670
CPUs, which support the AVX instruction set (see below), and using code
generated by the \textsf{gcc} compiler (version 4.8.2).

\subsection{Implementation details}
The FMM relations of Section~\ref{sec:prelim} and Appendix~\ref{app:math}
(using the rotation-accelerated M2L kernel of
Appendix~\ref{app:spherical:fast} when faster) have been implemented in
computer code. The code employs a one-sided version of the dual tree walk,
which considers the interactions $A\to B$ and $B\to A$ independently. The code
is written in the \texttt{C++} programming language and has been tested using
various compilers and hardware. The implementation employs vectorisation and
shared-memory parallelism as outlined below.

\subsubsection{Vectorisation}
Most current CPUs support vector sizes of 16 (SSE), 32 (AVX), or 64 (MIC)
bytes, allowing $K=2$, 4, or 8 identical simultaneous double-precision
floating-point operations (or twice as many in single precision). Because the
FMM kernels do not (usually) relate adjacent elements, their efficient
vectorisation is not straightforward (and well beyond compiler
optimisation). I explicitly implement a method computing $K$ M2L kernels
simultaneously. To this end, the multipole moments of the $K$ source cells are
loaded into a properly aligned buffer (similar to transposing a matrix)
before, and afterwards the $K$ field tensors are added from their
vector-buffer to the sink cells' field tensors. Unfortunately, this loading
and storing (which cannot be vectorised) reduces the speed-up obtained by the
simultaneous kernel computations.

Conversely, direct summation is perfectly suitable for vectorisation and a
speed-up of a factor $K$ is achievable. The code prefers direct summation
whenever this is deemed to be faster, based on a threshold for the number of
particle-particle interactions `caught' in a given cell-cell interaction.

\subsubsection{Multi-threading}
All parts of the implementation use multi-threading and benefit from multi-core
architectures. This is done via hierarchical task-based parallelism implemented
via \emph{threading building blocks} \citep[\textsf{tbb},][]{tbbBook}, an open
source task parallel library with a work-stealing scheduler. The algorithms for
multi-threaded tree building and dual tree walk are quite similar to those
described by \cite{TauraEtAl2012} and I refrain from giving details here.

\subsubsection{Precision and expansion order}
This study reports only on one particular implementation aimed at high
accuracy. It uses double precision (64 bits) floating-point arithmetic
throughout, $\vec{z}=\vec{z}_{\mathrm{ses}}$, and expansion orders $p\le20$.

\subsection{Wall-clock time versus accuracy}
\begin{figure}
  \begin{center}
    \resizebox{\columnwidth}{!}{\includegraphics{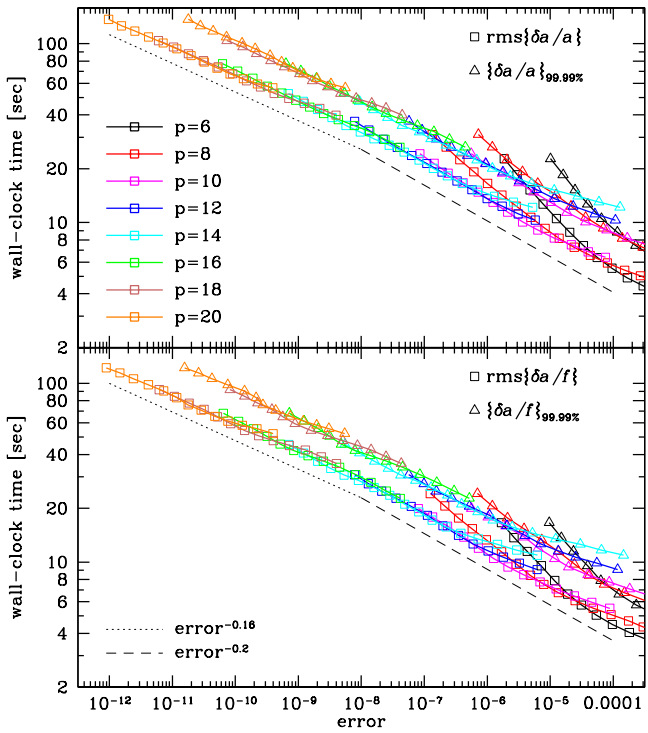}}
  \end{center}
  \caption{Wall-clock time versus relative (top) and scaled (bottom)
    acceleration error for $N=10^7$ particles drawn from a Plummer sphere.  The
    top panel reports runs using multipole-acceptance
    criterion~(\ref{eq:mac:better:a}) with a low-order-FMM estimate for $a_b$,
    while the bottom panel reports runs using multipole-acceptance
    criterion~(\ref{eq:mac:better:f}) with a low-order-FMM estimate for $f_b$.
    Each pair of open and filled symbols (of same colour and ordinate) refers to
    another FMM run with expansion order $p$ as indicated and a different value
    for parameter $\epsilon$ in equations~(\ref{eqs:mac:better}). The timings
    include all phases of the computation, including tree building and low-order
    estimation of $a_b$ or $f_b$---for comparison, the direct-summation
    calculation for obtaining the `true' accelerations took 25k seconds on
    the same hardware. The thin dotted and dashed lines are power laws with
    exponents $-0.16$ and $-0.2$, respectively.
    \label{fig:high:da:time}
  }
\end{figure}

I applied my implementation with criteria~(\ref{eq:mac:better:a}) and
(\ref{eq:mac:better:f}) to $N=10^7$ particles drawn from a Plummer sphere, and
using low-order estimates for $a_b$ and $f_b$ in
equations~(\ref{eqs:mac:better}). I varied the expansion order $p$ and the
accuracy parameter $\epsilon$ and for each run plot in
Fig.~\ref{fig:high:da:time} the total wall-clock time against the rms and the
99.99 percentile acceleration errors.

The rms error is always ten times smaller than the 99.99
percentile\footnote{The increase of this ratio to $\approx20$ towards small
  errors may well be caused by inaccuracies of the direct summation used for
  calculating the errors.}, implying the absence of extended large-error
tails. For any fixed expansion order $p$, the relation between time and
error can be approximated by a constant plus a power law that becomes flatter
for larger $p$. At any given error, there is an optimal expansion order $p$ in
the sense of providing the fastest approximation. When using this optimal
expansion order, the fastest FMM computation for a given error scales very
nearly like a power law with exponent $\sim-0.18$. Thus when reducing the
error by a factor ten, the computational costs rise only by a factor
$\sim1.5$.

Constraining the relative error (top panel of Fig.~\ref{fig:high:da:time}) is
slightly more costly than constraining the scaled error (bottom panel). This is
largely because $f>a$ as discussed in the caption to Fig.~\ref{fig:da:better},
but also because estimating $f$ is easier and faster than estimating
$\vec{a}$. Of course, the estimation of $\vec{a}$ can be easily avoided in
practice by using the accelerations from the previous time step.

\begin{figure}
  \begin{center}
    \resizebox{\columnwidth}{!}{\includegraphics{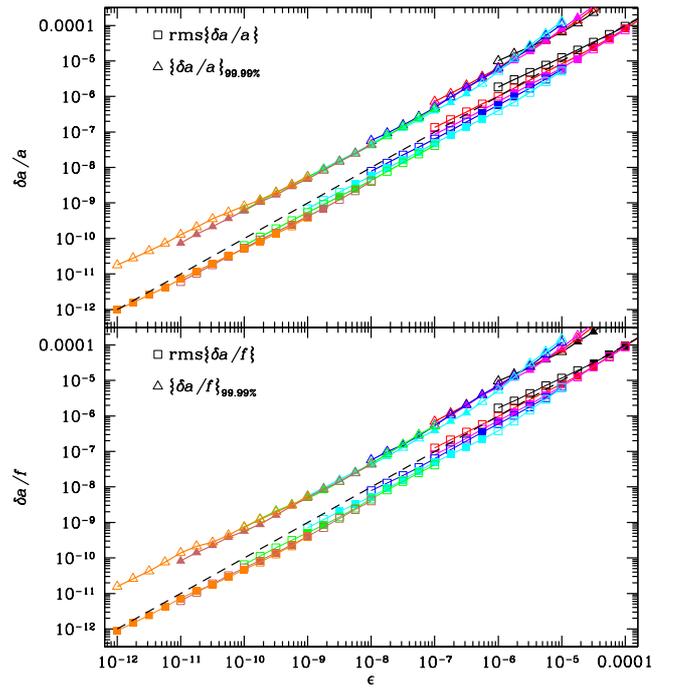}}
  \end{center}
  \caption{Rms value (squares) or 99.99 percentile (triangles) of the error
    $\delta a/a$ (top) or $\delta a/f$ (bottom) versus parameter $\epsilon$ for
    the same runs as in Fig.~\ref{fig:high:da:time} (using the same colour
    coding). Full symbols indicate that the expansion order is optimal,
    i.e.\ obtained minimal wall-clock time for the given error measure (lowest
    line in Fig.~\ref{fig:high:da:time}).
    \label{fig:eps:err}
  }
\end{figure}
\subsection{Accuracy versus parameter \boldmath $\epsilon$}
In any practical application there is, of course, no possibility to check on the
actual error, so it is important to test how well it is reflected by the
parameter $\epsilon$. As can be seen from Fig.~\ref{fig:eps:err}, the rms value
for the respective error ($\delta a/a$ if using criterion~\ref{eq:mac:better:a}
and $\delta a/f$ if using criterion~\ref{eq:mac:better:f}) is typically slightly
less than $\epsilon$ for the optimal expansion order $p$. At intermediate values
($\epsilon\sim10^{-8}$) the error is actually a factor $\sim2$ smaller. The
99.99 percentile of the errors is typically a factor ten larger.

\subsection{Complexity: scaling with the number \boldmath $N$ of particles}
The overall cost of my high-accuracy FMM implementation is dominated by the
computation of all node-node interactions during the dual tree walk. All other
phases (establishing the hierarchical tree structure, computing $\vec{z}$,
$\vec{s}$, and $\mathcal{M}_n^m$ for each cell; passing down $\mathcal{F}_n^m$
and evaluating gravity for each sink position) contribute much less (see
Tab.~\ref{tab:complexity}). When using a simple geometric multipole-acceptance
criterion, such as equation~(\ref{eq:mac:theta}), the FMM is well known to have
complexity $\mathcal{O}(N)$ \citep[e.g.][]{ChengGreengardRokhlin1999}. This is
because distant interactions contribute less than $\mathcal{O}(N)$, so that the
overall costs are dominated by the local interactions only \citep{Dehnen2002}.

\begin{table}
  \caption{Timings and errors as function of particle number $N$ for FMM runs
    with $p=10$, $\epsilon=10^{-6.25}$, and using multipole-acceptance
    criterion~(\ref{eq:mac:better:f}). The timings are given in seconds and
    refer to, respectively, the tree building; the estimation of $f$ via
    low-order FMM; the passing up of $\vec{z}$, $\vec{s}$, $\min\{f\}$, and
    $\mathcal{M}_n^m$; the dual tree walk; and the passing down of
    $\mathcal{F}_n^m$ and evaluation of gravity. See also
    Fig.~\ref{fig:complexity}.
    \label{tab:complexity}
  }
  \tabcolsep2pt
  \begin{tabular}{@{}rlllllll@{}}
    \multicolumn{1}{c}{$N$} &
    \multicolumn{1}{c}{$t_{\mathrm{tree}}$} &
    \multicolumn{1}{c}{$t_{\mathrm{ffmm}}$} &
    \multicolumn{1}{c}{$t_{\mathrm{up}}$} &
    \multicolumn{1}{c}{$t_{\mathrm{walk}}$} &
    \multicolumn{1}{c}{$t_{\mathrm{down}}$} &
    \multicolumn{1}{c}{$\{\delta a/\!\!f\}_{\mathrm{rms}}$} &
    \multicolumn{1}{c}{$\{\delta a/\!\!f\}_{99.99\%}$} \\
    \hline\\[-2ex]
    $10^4$ & 0.0004 & 0.0011 & 0.0012 & 0.0317 & 0.0005 &
    $2.81{\times}10^{-7}$ & $1.32{\times}10^{-6}$ \\
    $10^5$ & 0.0019 & 0.0042 & 0.0097 & 0.2131 & 0.0044 &
    $3.61{\times}10^{-7}$ & $2.51{\times}10^{-6}$ \\
    $10^6$ & 0.0290 & 0.0245 & 0.0957 & 1.5366 & 0.0431 &
    $3.85{\times}10^{-7}$ & $3.32{\times}10^{-6}$ \\
    $10^7$ & 0.2559 & 0.1897 & 0.8870 & 11.358 & 0.3999 &
    $4.05{\times}10^{-7}$ & $2.20{\times}10^{-6}$ \\
    \hline
  \end{tabular}
\end{table}

\begin{figure}
  \begin{center}
    \resizebox{83mm}{!}{\includegraphics{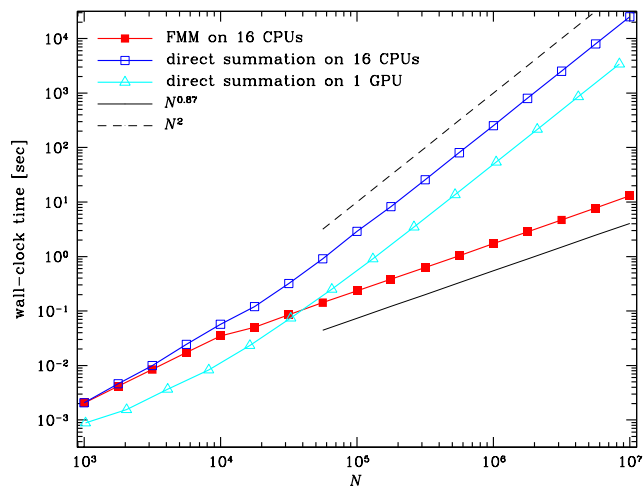}}
  \end{center}
  \caption{Wall clock time for the computation of the mutual gravitational
    forces between $N$ particles drawn from a Plummer sphere. The FMM (full
    squares) is parameterised (see Table~\ref{tab:complexity}) to yield
    acceleration errors very similar to those of direct summation on GPUs
    using the \textsf{sapporo} library (open triangles, using a
    \textsf{NVIDIA} K20M GPU
    accelerator)\protect\footnotemark[\getrefnumber{foot:GPU}]. The direct
    summation on 16 CPUs (open squares) uses double precision and besides the
    accelerations also computes the gravitational potential and the scale $f$
    (equation~\ref{eq:f}).}
  \label{fig:complexity}
\end{figure}

I am not aware of theoretical estimates for the complexity for the case of more
sophisticated multipole-acceptance criteria, but \cite{Dehnen2002} reports an
empirical scaling proportional to $N^{0.93}$ for his approach of a
mass-dependent opening angle. Table~\ref{tab:complexity} and
Fig.~\ref{fig:complexity} present the timings obtained with my implementation
using $p=10$, $\epsilon=10^{-6.25}$, and low-order FMM estimates of $f_b$ in
equation~(\ref{eq:mac:better:f}). With these settings, the acceleration errors
are comparable to those generated via the \textsf{sapporo} library on a GPU (the
current state-of-the-art force solver for collisional $N$-body simulations), as
reported in Section~\ref{sec:error:dist:direct}.

From Table~\ref{tab:complexity}, it can be seen that the costs for tree building
grow faster than linearly with $N$ ($N\ln N$ is expected), those for the upward
and downward passes roughly linearly with $N$ (as expected), but those for the
FMM estimation of $f$ and the dual tree walk less than linearly. As a result,
the total computational costs are very well fit by the power law $N^{0.87}$ for
$N>10^4$, see Fig.~\ref{fig:complexity}.

This Figure also shows the timings for a (double-pre\-cision) direct-summation
on the same hardware (yielding much more accurate accelerations) and for a
mixed-precision direct-summation on a GPU using the \textsf{sapporo} library
(yielding comparably accurate accelerations)\footnote{\label{foot:GPU}The
  timings for the sapporo library also include additional computations (nearest
  neighbour finding and neighbour listing). These contribute negligibly at large
  $N$, but at small $N$ they are, together with latency on the GPU, responsible
  for the deviation of the observed complexity from $N^2$.}. At large (but
realistic) $N$ FMM out-performs direct summation, even if accelerated using a
GPU.

\subsection{Scaling with the number of CPUs}
\begin{figure}
  \begin{center}
    \resizebox{70mm}{!}{\includegraphics{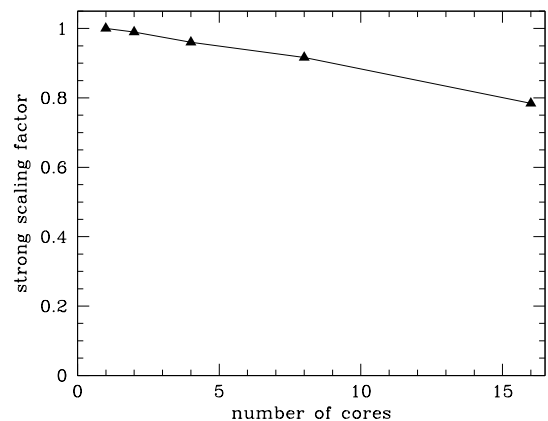}}
  \end{center}
  \caption{Strong scaling of my implementation for the computation of the mutual
    forces for $N=10^7$ particles with $p=10$ and $\epsilon=10^{-6.25}$.
    \label{fig:scaling} }
\end{figure}

Fig.~\ref{fig:scaling} plots the strong scaling factor $t_1/nt_n$ for my
multi-threaded implementation. The scaling drops to 80\% for 16 cores, which is
not untypical for multi-threaded programs. This drop is presumably caused by
imbalances at synchronisation points, of which the implementation has many. Most
of these are not algorithmically required, but allow for a much easier
implementation. Clearly, any massively parallel implementation needs to address
this issue to retain good scaling for large numbers of processors.

\section{Beyond simple gravity approximation}
\label{sec:beyond}
So far, I have considered the approximate computation of the unsoftened
gravitational potential and acceleration at all particle positions with equal
relative (or scaled) accuracy. However, the fast multipole method can be easily
modified or extended beyond that.

For example, one may want to have individual accuracy parameters $\epsilon_b$
instead of a global one. This is easily accommodated by replacing
$\epsilon\min_{b\in B}\{a_b\}$ in criterion~(\ref{eq:mac:better:a}) with
$\min_{b\in B}\{\epsilon_ba_b\}$ and analogously for
criterion~(\ref{eq:mac:better:f}).

When using individual $\epsilon_b$, but also in general, it may be beneficial
to adapt the expansion order $p$ to the accuracy actually required for a given
cell$\,\to\,$cell interaction. This could be implemented by using the lowest
$p\le p_{\max}$ for which the multipole-acceptance criterion is satisfied.

\subsection{Force computation for a subset of particles}
Most $N$-body codes employ adaptive individual time steps for each
particle. The standard technique is \citeauthor{Makino1991}'s
(\citeyear{Makino1991}) block-step scheme, where the forces of all
\emph{active} particles are computed synchronously. Active are those particles
with time step smaller than some threshold (which varies from one force
computation to the next).

\begin{figure}
  \begin{center}
    \resizebox{83mm}{!}{\includegraphics{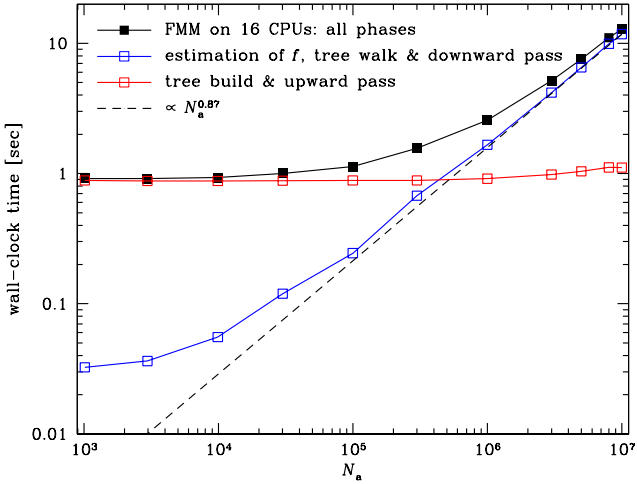}}
  \end{center}
  \caption{Wall clock time for the computation of gravity for the innermost
    $N_{\mathrm{a}}$ of $N=10^7$ particles (of a Plummer sphere), using the
    same parameters ($p$ and $\epsilon$) as in Fig.~\ref{fig:complexity}.}
  \label{fig:subset}
\end{figure}
When using FMM in such a situation, only interactions with sink cells contain
at least one active particle must be considered. If the fraction of active
particles in such cells is small (but non-zero), FMM becomes much less
efficient per force computation. Fortunately, however, active particles are
typically spatially correlated (because the time steps of adjacent particles
are similar), such that the fraction of active particles is either zero or
large.

I performed some practical tests, where only particles within some distance
from the origin of the system were considered active. Fig.~\ref{fig:subset}
plots the wall-clock time vs.\ the number $N_{\mathrm{a}}$ of active particles
for $N=10^7$. As expected the costs for the preparation phase (tree building and
upward pass) are largely independent of $N_{\mathrm{a}}$ (the slight increase
of the red curve at large $N_{\mathrm{a}}$ is because $\vec{s}$ and
$\rho_{\vec{s}}$ are computed as part of the upward pass, but only for cells
with active particles).

The costs for the interaction and downward pass, on the other hand, decrease
roughly like $N_{\mathrm{a}}^{0.87}$ for $N_{\mathrm{a}}\gtrsim10^4$. The net
effect is that for $N_{\mathrm{a}}/N\lesssim0.01$, the costs are almost
completely dominated by the preparation phase, and hence independent of
$N_{\mathrm{a}}$. The precise point of this transition depends on $N$ and the
FMM parameters. For smaller $N$ and/or more accurate forces, the relative
contribution of the tree-walk phase increases and the transition occurs at
smaller $N_{\mathrm{a}}$.

There is certainly some room for improvement by, e.g.\ using a smaller
expansion order $p$ than is optimal for $N_{\mathrm{a}}=N$ and/or re-cycling
the tree structure from the previous time step. Both measures reduce the costs
of the preparation phase and increase that of the interaction phase (at given
$\epsilon$), but shall reduce the overall costs if $N_{\mathrm{a}}\ll N$.
\subsection{Softened gravity or far-field force}
Gravitational softening amounts to replacing the Newtonian Greens function
$\psi=|\vec{r}|^{-1}$ by \citep{Dehnen2001}
\begin{equation} \label{eq:greens:soft}
  \psi(\vec{r}) = h^{-1} \varphi(|\vec{r}|/h)
\end{equation}
with softening length $h$ and \emph{softening kernel} $\varphi(q)\to q^{-1}$
as $q\to\infty$. This corresponds to replacing each source point by a smooth
mass distribution with density $\mu_b\,\varrho(\vec{x}-\vec{x}_b)$, where
\begin{equation}
  \varrho(\vec{r}) = - (4\pi)^{-1} \vec{\nabla}^2 \psi(\vec{r}).
\end{equation}
This Greens function~(\ref{eq:greens:soft}) is no longer harmonic and harmonic
FMM cannot be used. One obvious option is to use the more general Cartesian FMM
of Appendix \ref{app:Cartesian} \citep{Dehnen2002}. The computational costs of
this approach grow faster with expansion order $p$, such that small
approximation errors (requiring high $p$) become significantly more
expensive. However, small approximation errors are hardly required in situations
where gravitational softening is employed. Alternatively, if softening is
restricted to a finite region, i.e.\ if $\varrho(\vec{r})=0$ for $|\vec{r}|\ge
h$, harmonic FMM can still be used to compute gravity from all sources at
distances $|\vec{r}|\ge h$, while direct summation could be used for
\emph{neighbours}, sources at $|\vec{r}|<h$. This approach is sensible only if
the number of neighbours is sufficiently bounded (so that the cost incurred by
the direction summation remains small). This is the case, in particular, if the
number of neighbours is kept (nearly) constant by adapting the individual
softening lengths $h_i$ in order to adapt the numerical resolution
\citep{PriceMonaghan2007}.

In practice, this requires to carry with each cell the radius
$h_{\vec{z}}>\rho_{\vec{z}}$ of the smallest sphere centred on $\vec{z}$ which
contains all softening spheres of its sources, and allow a FMM interaction $A\to
B$ only if $|\vec{z}_{A}-\vec{s}_{B}|>h_{\vec{z},A}+\rho_{\vec{s},B}$.

The same technique can be used to restrict the FMM approximation to the far
field for each particle, i.e.\ the force generated by all sources outside of a
sphere of known radius $h_b$ around $\vec{x}_b$.

\subsection{Jerk, snap, crackle, and pop}
The jerk is the total time derivative of the acceleration
\begin{equation}
  \vec{j} = \dot{\vec{a}} =
  \frac{\mathrm{d}}{\mathrm{d}t}
  \frac{\p \Psi}{\p\vec{x}}.
\end{equation}
The simplest way to estimate this using FMM, is to not allow the expansion
centres to have any velocity ($\dot{\vec{z}}=\dot{\vec{s}}=0$), such that
differentiating the FMM relations~(\ref{eqs:fmm}) w.r.t.\ time gives
\begin{subequations}
  \label{eqs:fmm:dot}
  \begin{eqnarray}
    \label{eq:L2P:dot}
    \dot{\Psi}_{n}^m(\vec{x}_b) &=& \sum_{k=0}^{p-n}\,\sum_{l=-k}^k\,\left[
      \Upsilon_k^{l\ast}(\vec{s}_{B}-\vec{x}_b)\,
      \dot{\mathcal{F}}_{n+k}^{m+l}(\vec{s}_{B})
      + \right.\nonumber \\ && \quad \left.
      \dot{\Upsilon}_k^{l\ast}(\vec{s}_{B}-\vec{x}_b,-\dot{\vec{x}}_b)\,
      \mathcal{F}_{n+k}^{m+l}(\vec{s}_{B})\right] + \delta\dot{\Psi}_{n}^m,
    \\
    \label{eq:M2L:dot}
    \dot{\mathcal{F}}_n^{m}(\vec{s}_{B}) &=&
    \sum_{k=0}^{p-n}\sum_{l=-k}^{k}\,\dot{\mathcal{M}}_k^{l\ast}(\vec{z}_{A})\,
    \Theta_{n+k}^{m+l}(\vec{s}_{B}-\vec{z}_{A}),
    \\
    \label{eq:P2M:dot}
    \dot{\mathcal{M}}_n^m(\vec{z}_{A}) &=& \sum_{a\in A} \mu_a\,
    \dot{\Upsilon}_n^m(\vec{x}_a-\vec{z}_{A},\dot{\vec{x}}_a).
  \end{eqnarray}
\end{subequations}
and the jerk follows from $\vec{j}=-(\Re\{\dot{\Psi}_1^1\},
\Im\{\dot{\Psi}_1^1\},\dot{\Psi}_1^0)$.  Since $\dot{\vec{z}}=\dot{\vec{s}}=0$,
the M2M and L2L kernels (equations \ref{eq:M2M} and \ref{eq:L2L}) work also for
the time derivatives $\dot{\mathcal{M}}_n^m$ and $\dot{\mathcal{F}}_n^{m}$ of
the multipoles and field tensors, respectively. The relations for the next
order, the snap $\vec{s}=\ddot{\vec{a}}$, can be derived by differentiating yet
again.

With each additional order (jerk, snap, crackle, pop, $\dots$), the
computational cost of the combined M2L kernels is not more than the
corresponding multiple of the ordinary M2L kernel (i.e.\ acceleration plus jerk
are twice as costly as just acceleration). This is a direct consequence of not
allowing cell-centre velocities hence preventing the terms depending on
$\vec{z}$ or $\vec{s}$ in equations~(\ref{eqs:fmm}) to carry any time
dependence. In contrast, the computational costs of the P2M and L2P kernels
grows quadratically with the order of time derivative. This is not really a
problem, since those kernels are only needed once per particle, while the M2L
kernel is typically used $\gtrsim100$ times more often.

\subsection{The tidal field}
FMM can also be used to approximate the Hessian
\begin{equation}
   \tens{T}_b = \vec{\nabla}\vec{\nabla}\Psi(\vec{x}_b)
\end{equation}
of the potential, which is given by the components of $\Psi_2^m$ via
\begin{equation}
  \begin{array}{lccll}
    \mathsf{T}_{xx} &=&-& \tfrac{1}{2}\Psi_2^0+\tfrac{1}{2}\Re\{\Psi_2^2\}, &
    \\[0.5ex]
    \mathsf{T}_{xy} &=&-& \tfrac{1}{2}\Im\{\Psi_2^2\}, & \\[0.5ex]
    \mathsf{T}_{xz} &=& & \Re\{\Psi_2^1\}, & \\[0.5ex]
    \mathsf{T}_{yy} &=&-& \tfrac{1}{2}\Psi_2^0-\tfrac{1}{2}\Re\{\Psi_2^2\}, &
    \\[0.5ex]
    \mathsf{T}_{yz} &=& & \Im\{\Psi_2^1\}, & \text{and} \\[0.5ex]
    \mathsf{T}_{zz} &=& & \Psi_2^0
    \end{array}
\end{equation}
(in particular $\mathrm{tr}(\tens{T})=0$ as expected). Note, however, that
the accuracy of this approximation is lower than that for the acceleration.
$\tens{T}$ is of particular interest in collisionless $N$-body modelling,
when
\begin{equation}
  \tau_b \le \eta\,||\tens{T}_b||^{-1/4}
\end{equation}
with dimensionless parameter $\eta\ll1$ has been suggested as criterion for
individual particle time steps \citep{DehnenRead2011}. The matrix norm of
$\tens{T}$ may be directly computed from $\Psi_2^m$ as
\begin{equation}
  ||\tens{T}_b|| = 
  \tfrac{3}{2}\left|\Psi_2^0\right|^2 +
  2\left|\Psi_2^1\right|^2 +
  \tfrac{1}{2} \left|\Psi_2^2\right|^2.
\end{equation}

\section{Discussion and Conclusions}
\label{sec:conclude}
The fast multipole method (FMM) approximates the computation of the mutual
forces between $N$ particles. I have derived the relevant mathematical
background, giving much simpler formul\ae\ than the existing literature, for the
case of unsoftened gravity, when the harmony of the Greens function allows
significant reduction of the computational complexity.

Like the tree code, my FMM implementation uses a hierarchical tree of spatial
cells. Unlike the tree code, FMM uses cell$\,\to\,$cell interactions, which
account for all interactions between sources in the first cell and sinks in the
second. Almost all distant particle$\,\to\,$particle interactions are `caught'
by fewer than $\mathcal{O}(N)$ cell$\,\to\,$cell interactions, such that local
interactions, requiring $\mathcal{O}(N)$ computations, dominate the overall
workload \citep{Dehnen2002}. With the tree code, the situation is reversed: the
distant interactions require $\mathcal{O}(N\ln N)$ computations and dominate the
overall work. This implies that FMM has the best complexity of all known force
solvers. What is more, the predominance of local as opposed to distant
interactions makes FMM ideally suited for applications on super-computers, where
communications (required by distant interactions) are increasingly more costly
than computations. However, FMM is inherently difficult to parallelise and this
study considered only a multi-threaded implementation with a task-parallel dual
tree walk (the core of FMM).

Most previous implementations of FMM considered simple choices for the cell's
multipole- and force-expansion centres and the multipole-acceptance criterion
which decides whether a given cell$\,\to\,$cell interaction shall be processed
via the multipole expansion or be split into daughter
interactions. Traditionally, a simple opening-angle based multipole-acceptance
criterion has been used and cell centres equal to either the cell's geometric
centre or its centre of mass. These choices, which presumably were based on
computational convenience and intuition, inevitably result in a wide
distribution of individual relative force errors with extended tails reaching
$\sim1000$ times the median.

The main goal of this study was avoid such extended tails of large force errors
and to minimise the computational effort at a given force accuracy. The key for
achieving this goal is a reasonably accurate estimate, based on the multipole
power of the source cell and the size of the sink cell, for the actual force
error incurred by individual cell$\,\to\,$cell interactions. Based on the
insight from this estimate, I set the cell's force-expansion centres to (an
approximation of) the centre of the smallest sphere enclosing all its particles,
when the cell size and hence the error estimates are minimal. I also use the
new estimates in the multipole-acceptance criterion, such that each
cell$\,\to\,$cell interaction is considered on the merit of the error it likely
incurs. This results in very well behaved distributions of the relative force
errors, provided an initial estimate for the forces is at hand. This can either
be taken from the previous time step or obtained via low-accuracy FMM.

After these improvements, the method has only two free parameters: the
expansion order $p$ and a parameter $\epsilon$ for the relative force
error. Experiments showed that the actual rms relative force error is
typically somewhat less than $\epsilon$, while for any given $\epsilon$ there
is an optimum $p$ at which the computational cost are minimal. For
$\epsilon=10^{-6.25}$, for example, $p=10$ is optimal and the accelerations
errors are comparable to those of direct summation on a GPU (the current
state-of-the-art method for collisional $N$-body simulations). With these
parameter settings, the computational costs scale like $N^{0.87}$ for large
$N$ and the method out-performs any direct-summation implementation for
$N\gtrsim10^5$. When computing only the forces for $N_{\mathrm{a}}<N$ of $N$
particles, the costs are roughly proportional to $N_{\mathrm{a}}^{0.87}$ for
$N_{\mathrm{a}}/N\gtrsim0.01$, but become independent of $N_{\mathrm{a}}$
below that (where the costs for tree building dominate). For large $N$, this
is still significantly faster than direct summation.

An implementation of the FMM on a GPU accelerator should yield a further
significant speed-up compared to my CPU-based implementation, though this is
certainly a challenging task, given that FMM is algorithmically more complex
than direct summation or a tree code (both of which have been successfully
ported to the GPU). Presumably a somewhat lesser challenge is a massively
parallel implementation of the method, which can be run on a super computer.

A practical application of FMM in an actual collisional $N$-body simulation
would be very interesting. Since the force between close neighbours is always
computed directly (in double precision) as explained earlier, close encounters
can be treated essentially in the same fashion as with existing techniques.
However, an unfortunate hindrance to an application of the presented
techniques originates from the long marriage of existing collisional $N$-body
techniques with direct summation. Methods, such as the Ahmad-Cohen neighbour
scheme, to reduce the need for the costly far-field force summations are not
necessary with FMM, and the existing $N$-body tools are not well suited for an
immediate application of FMM.

\begin{acknowledgements}
The author thanks Joachim Stadel for many helpful discussions and the suggestion
to allow $\vec{z}\neq\vec{s}$, Alessia Gualandris for running \textsf{sapporo}
to provide the data for Fig.~\ref{fig:da:gpu}, and Simon Portegies Zwart and
Jeroen B\'edorf for providing the timings for \textsf{sapporo\,2} in
Fig.~\ref{fig:complexity}. This work was supported by STFC consolidated grant
ST/K001000/1.
\end{acknowledgements}
\bibliographystyle{spbasic}
{\small
\bibliography{refs}
}
\appendix
\section{Derivation of the FMM relations} \label{app:fmm}
Here, the FMM relations given in Section~\ref{sec:prelim} are derived and
motivated. Differently from the main text, the multipole and force expansion
centres, $\vec{z}$ and $\vec{s}$, are not explicitly distinguished and instead
$\vec{z}$ is used for either. The general case $\vec{z}\neq\vec{s}$ is a trivial
generalisation.
\label{app:math}
\begin{figure}
  \centerline{\resizebox{70mm}{!}{\includegraphics{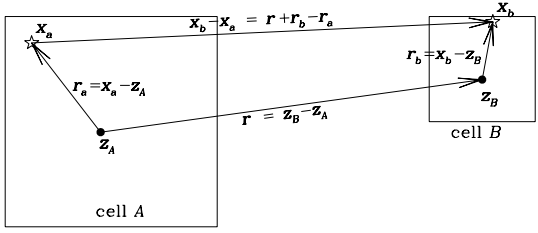}}}
  \caption{The Greens function for the interaction between particles $a$ and $b$
    at positions $\vec{x}_a$ and $\vec{x}_b$ and residing in cells $A$ and $B$,
    is approximated by Taylor expanding the Greens function about $\vec{r}$, the
    distance vector between the cell centres $\vec{z}_A$ and $\vec{z}_B$.
    \label{fig:fmm}}
\end{figure}
\subsection{Cartesian FMM}
\label{app:Cartesian}
The distance vector $\vec{x}_b - \vec{x}_a$ between two particles residing in
two well-separated cells $A$ and $B$, respectively, can be decomposed into three
components (see also Fig.~\ref{fig:fmm})
\begin{equation}
  \vec{x}_b - \vec{x}_a = \vec{r} + \vec{r}_b - \vec{r}_a
\end{equation}
with $\vec{r}_a\equiv\vec{x}_a-\vec{z}_A$, $\vec{r}_b\equiv\vec{x}_b-\vec{z}_B$,
and $\vec{r}\equiv\vec{z}_B-\vec{z}_A$. The Taylor expansion of the general
Greens function $\psi(\vec{x}_b - \vec{x}_a)$ in $\vec{r}_a$ \emph{and}
$\vec{r}_b$ up to order $p$ then reads\footnote{Using \emph{multi-index}
  notation $\vec{n}\equiv(\mathrm{n}_x,\mathrm{n}_y,\mathrm{n}_z)$ with
  $n\equiv |\vec{n}| \equiv \mathrm{n}_x+\mathrm{n}_y+\mathrm{n}_z$, such that
  the first sum in~(\ref{eq:fmm:greens}) is over non-negative integer triples
  $\vec{n}$ with $\mathrm{n}_x+\mathrm{n}_y+\mathrm{n}_z \le p$. Furthermore
  $\vec{n}! \equiv \mathrm{n}_x!\;\mathrm{n}_y!\;\mathrm{n}_z!$ and
  $\vec{r}^{\vec{n}} \equiv
  r_x^{\mathrm{n}_x}\;r_y^{\mathrm{n}_y}\;r_z^{\mathrm{n}_z}$.}
\begin{equation} \label{eq:fmm:greens}
  \psi(\vec{x}_b-\vec{x}_a) =
  \sum_{|\vec{n}|=0}^p\sum_{|\vec{m}|=0}^{p-|\vec{n}|}
  \frac{(-1)^{|\vec{m}|}}{\vec{n}!\vec{m}!}\,
  \vec{r}_b^{\vec{n}}\,\vec{r}_a^{\vec{m}}\, \vec{\nabla}^{\vec{n}+\vec{m}}
  \psi(\vec{r}) + \mathcal{R}_p .
\end{equation}
This series converges (the remainder $\mathcal{R}_p\to0$) as $p\to\infty$, if
$|\vec{r}_a+\vec{r}_b|<|\vec{r}|$. Inserting (\ref{eq:fmm:greens}) into the
expression
\begin{equation}
  \Psi_{A\to B}(\vec{x}_b) = \sum_{a\in A}\mu_a\, \psi(\vec{x}_b-\vec{x}_a)
\end{equation}
for the (negative) potential due to all source points in cell $A$ and for any
sink position $\vec{x}_b$ in cell $B$, one obtains after re-arranging
\begin{subequations} \label{eqs:fmm:cartesian}
  \begin{eqnarray}
    \label{eq:evaluate}
    \Psi_{A\to B}(\vec{x}_b) &=& \sum_{|\vec{n}|=0}^p\frac{1}{\vec{n}!}\,
    (\vec{x}_b-\vec{z}_B)^{\vec{n}}\,\mathsf{F}_{\vec{n}}(\vec{z}_B) +
    \delta\Psi_{A\to B}, \\
    \label{eq:fieldtensor}
    \mathsf{F}_{\vec{n}}(\vec{z}_B) &\equiv& \sum_{|\vec{m}|=0}^{p-|\vec{n}|}
    \mathsf{M}_{\vec{m}}(\vec{z}_A)\,
    \mathsf{D}_{\vec{n}+\vec{m}}(\vec{z}_B-\vec{z}_A), \\
    \label{eq:multipole}
    \mathsf{M}_{\vec{m}}(\vec{z}_A) &\equiv& \sum_{a\in A}
    \mu_a\,\frac{(-1)^m}{\vec{m}!}  \,(\vec{x}_a-\vec{z}_A)^{\vec{m}}
  \end{eqnarray}
  with the derivatives $\mathsf{D}_{\vec{n}}(\vec{r}) \equiv
  \vec{\nabla}^{\vec{n}} \psi(\vec{r})$.  The FMM algorithm essentially works
  these equations backwards: in a first step, the \emph{multipoles}
  $\mathsf{M}_{\vec{m}}(\vec{z})$ are computed for each cell via
  (\ref{eq:multipole}) and by utilising those of daughter cells via the shifting
  formula
  \begin{equation} \label{eq:M:shift}
    \mathsf{M}_{\vec{m}}(\vec{z}+\vec{x}) =
    \sum_{\vec{n}=0}^{\vec{m}} \frac{\vec{x}^{\vec{n}}}{\vec{n}!}
    \mathsf{M}_{\vec{m}-\vec{n}}(\vec{z}).
  \end{equation}
  Second, for each cell the \emph{field tensors}
  $\mathsf{F}_{\vec{n}}(\vec{z})$ of all its interactions are computed via
  (\ref{eq:fieldtensor}) and added up. Finally, the field tensors are passed
  down the tree, utilising the shifting formula
  \begin{equation} \label{eq:F:shift}
    \mathsf{F}_{\vec{n}}(\vec{z}+\vec{x}) =
    \sum_{|\vec{m}|=0}^{p-|\vec{n}|} \frac{\vec{x}^{\vec{m}}}{\vec{m}!}
    \mathsf{F}_{\vec{n}+\vec{m}}(\vec{z}),
  \end{equation}
\end{subequations}
and the potential (and its derivative, the acceleration) is evaluated via
(\ref{eq:evaluate}) at each sink position. Equations (\ref{eqs:fmm:cartesian})
are the basis of Cartesian FMM, such as implemented in \citeauthor{Dehnen2002}'s
(\citeyear{Dehnen2000:falcON,Dehnen2002}) \textsf{falcON} algorithm.

At each order $n=|\vec{n}|$, there are $\binom{n+2}{2}$ coefficients
$\mathsf{F}_{\vec{n}}$ (as well as $\mathsf{M}_{\vec{n}}$ and
$\mathsf{D}_{\vec{n}}$), and the total number of coefficients up to order $p$
is $\binom{p+3}{3}$. The computational effort of the resulting algorithm is
dominated by their computation in (\ref{eq:fieldtensor}), which requires about
$\binom{p+6}{6}$ multiplications. Thus at large $p$ a straightforward
application of this method approaches an operation count of
$\mathcal{O}(p^6)$. The computation (\ref{eq:fieldtensor}) of the field tensors
is essentially a convolution in index space and hence can be accelerated using a
fast Fourier technique with costs $\mathcal{O}(p^3\ln p)$ (but see
footnote~\ref{foot:O}).

\subsection{Harmonic tensors}
\label{app:Harmonic}
For the important case $\psi=|\vec{r}|^{-1}$ corresponding to gravitational and
electrostatic forces, the above method can be improved by exploiting that this
Greens function is \emph{harmonic}, i.e.\ $\vec{\nabla}^2\psi= 0$ for
$|\vec{r}|>0$. As a consequence, the $\mathsf{D}_{\vec{n}} =
\vec{\nabla}^{\vec{n}} \psi$ are harmonic too and satisfy
\begin{equation} \label{eq:traceless}
  \vec{\nabla}^2\mathsf{D}_{\vec{k}} = 
  \mathsf{D}_{\vec{k}+(2,0,0)} +
  \mathsf{D}_{\vec{k}+(0,2,0)} +
  \mathsf{D}_{\vec{k}+(0,0,2)} = 0.
\end{equation}
In other words: $\mathsf{D}_{\vec{n}}$ is \emph{traceless}. At given degree
$n=k+2$, equation~(\ref{eq:traceless}) gives $\binom{n}{2}$ constraints such
that of the $\binom{n+2}{2}$ terms only $2n+1$ are truly independent. In inner
products, a traceless tensor only `sees' the traceless part of its co-operand:
\begin{equation} \label{eq:innerprod}
  \sum_{|\vec{n}|=n} \ol{\mathsf{A}}_{\vec{n}} \mathsf{B}_{\vec{n}}  =
  \sum_{|\vec{n}|=n} \ol{\mathsf{A}}_{\vec{n}} \ol{\mathsf{B}}_{\vec{n}}  =
  \sum_{|\vec{n}|=n} \mathsf{A}_{\vec{n}} \ol{\mathsf{B}}_{\vec{n}},
\end{equation}
where the `reduced' tensor $\ol{\mathsf{A}}_{\vec{n}}$ denotes the traceless
part of $\mathsf{A}_{\vec{n}}$. Furthermore, $\ol{\vec{r}^{\vec{n}}}$ is
related to $\mathsf{D}_{\vec{n}}$ via
\begin{equation} \label{eq:r:tr}
  \mathsf{D}_{\vec{n}}(r) =
  (-1)^{n}\,(2n-1)!!\,\frac{\ol{\vec{r}^{\vec{n}}}}{r^{2n+1}}.
\end{equation}
With these relations, the Taylor series of the harmonic Greens function becomes,
for $r>x$ 
\begin{equation} \label{eq:Cartesian}
  \frac{1}{|\vec{r}-\vec{x}|}
  = \sum_{n=0}^\infty
  \frac{x^n}{r^{n+1}}
  \sum_{|\vec{n}|=n}\frac{(2n-1)!!}{\vec{n}!}\;
  \ol{\hat{\vec{x}}^{\vec{n}}}\, \ol{\hat{\vec{r}}^{\vec{n}}},
\end{equation}
which is the Cartesian equivalent to the spherical harmonic expansion
\begin{equation} \label{eq:spherical}
  \frac{1}{|\vec{r}-\vec{x}|} = \sum_{n=0}^\infty
  \frac{x^n}{r^{n+1}} \sum_{m=-n}^n
  Y_n^{m}(\hat{\vec{x}})\,
  Y_n^{m\ast}(\hat{\vec{r}})
\end{equation}
(see eq.~(\ref{eq:Ynm:def}) for a definition of $Y_n^m$).  While at each order
$n$ there are only $2n+1$ truly independent terms, the expansion
(\ref{eq:Cartesian}) still carries all $\binom{n+2}{2}$ terms, amounting to a
total of $\binom{p+3}{3}$ terms in an expansion up to order $p$.  The equivalent
spherical harmonic expansion (\ref{eq:spherical}) only carries $2n+1$ terms per
order\footnote{In equation (\ref{eq:spherical}), the $Y_l^m$ are complex-valued
  for $m\neq0$, but because of their symmetry $Y_l^{m\ast}=(-1)^mY_l^{-m}$ there
  are only $2n+1$ independent real-valued components per order $n$.} amounting
to a total of $(p+1)^2$, i.e.\ at large $p$ is much preferable.

The number of terms actually used can be reduced to $(2n+1)$ per order, for
example, by omitting all terms with $\mathsf{n}_z>1$ and recover their
contribution via recursive application of
\begin{equation} \label{eq:Dz:DxDy}
  \mathsf{D}_{\vec{k}+(0,0,2)} = -
  \mathsf{D}_{\vec{k}+(2,0,0)} - \mathsf{D}_{\vec{k}+(0,2,0)}
\end{equation}
\citep{Applequist1989,HinsenFelderhof1992}. However, the resulting algebraic
challenges are considerable, though the overall computational effort could
well be reduced to $\mathcal{O}(p^3)$ operations (Joachim Stadel, private
communication), but I am not aware of a systematic demonstration.

\subsection{Spherical harmonics}
\label{app:spherical:complex}
The algebraic complications with obtaining an efficient Cartesian FMM stem from
the fact that the Laplace operator involves three terms, such that the resulting
recovery relation~(\ref{eq:Dz:DxDy}) has two terms instead of one on the
right-hand side. This problem can be avoided by Taylor expanding in other than
Cartesian coordinates where the Laplace operator involves only two instead of
three terms.

The simplest possibility is a linear combination of Cartesian coordinates with
complex coefficients. The standard FMM relations emerge from replacing $x$ and
$y$ with
\begin{equation} \label{eq:xi:eta}
  \xi\equiv \tfrac{1}{2}(x+iy) \qquad\text{and}\qquad \eta\equiv
  -\tfrac{1}{2}(x-iy)=-\xi^\ast,
\end{equation}
while keeping $z$. Then $\p_\xi=\p_x-i\p_y$ and $\p_\eta=-\p_x-i\p_y$, such
that $\p^2_x+\p^2_y=-\p_\xi\p_\eta $ and hence for harmonic
functions
\begin{equation} \label{eq:dd}
  \p_\xi\,\p_\eta=\p_z^2,
\end{equation}
or $\mathsf{D}_{\vec{k}+(0,0,2)}=\mathsf{D}_{\vec{k}+(1,1,0)}$ in place of
equation~(\ref{eq:Dz:DxDy}). With this relation one can eliminate all mixed
$\xi$-$\eta$ derivatives in favour of $z$ derivatives. This in turn allows a
reduction in the number of indices from three to two by using the total number
$n$ of derivatives and the number $|m|$ of $\xi$ (for $m<0$) or $\eta$
derivatives (for $m>0$).

Somewhat surprisingly, the relations required for FMM are hardly covered by the
rich literature on spherical harmonics (and FMM). To derive the relevant
formul\ae, I follow the ideas of \citeauthor{Maxwell1892}
(\citeyear{Maxwell1892}, see also \citealt{James1969}) and define the
differential operator
\begin{equation} \label{eq:Delta}
  \Delta_n^m \equiv \left\{
  \begin{array}{lr}
    (-\p_z)^{n-|m|}\;\p_\xi ^{|m|} &\quad -n \le m < 0,\\[0.5ex]
    (-\p_z)^{n-|m|}\;\p_\eta^{|m|} &       0 \le m \le n,\\[0.5ex]
    0 & |m| > n.
  \end{array}
  \right. \\
\end{equation}
When applied to harmonic functions, this operator satisfies
\begin{equation} \label{eq:Delta:prod}
  \Delta_n^m\,\Delta_k^l = \Delta_{n+k}^{l+m},
\end{equation}
which can be shown via equation~(\ref{eq:dd}) and is inevitably linked to
\begin{equation} \label{eq:sym}
  \Delta_n^{-m}=(-1)^m\Delta_n^{m\ast}.
\end{equation}
Since $\psi=|\vec{r}|^{-1}$ is harmonic, its derivatives
\begin{equation} \label{eq:The:def}
  \Theta_n^m(\vec{r}) \equiv \Delta_n^m\,|\vec{r}|^{-1}
\end{equation}
are harmonic too. Moreover, the functions $\Theta_n^m(\vec{r})$ are homogeneous
of degree $-(n+1)$, i.e.\ $\Theta_n^m(\alpha\vec{r}) = \alpha^{-(n+1)}
\Theta_n^m(\vec{r})$. I also define the \emph{solid spherical harmonic} of
degree $n$ as
\begin{equation} \label{eq:Ups:def}
  \Upsilon_n^m(\vec{r}) \equiv
  	\frac{r^{-1}\,\Theta_n^m(\vec{r}/r^2)}{(n-m)!\,(n+m)!} =
  	\frac{r^{2n+1}\,\Theta_n^m(\vec{r})}{(n-m)!\,(n+m)!}.
\end{equation}
That $\Upsilon_n^m$ is harmonic follows from the fact that if $f(\vec{r})$ is
harmonic, then so is $r^{-1}f(\vec{r}/r^2)$ \citep[][try this with your
  undergraduate students]{Hobson1931}. Note that $\Upsilon_n^m(\vec{r})$ is just
a homogeneous polynomial of total degree $n$ in $x$, $y$ and $z$. These
harmonics are related to the usual normalised \emph{surface spherical harmonic}
\begin{equation} \label{eq:Ynm:def}
  Y_n^m(\hat{\vec{r}}) \equiv (-1)^m\,\sqrt{\frac{(n-m)!}{(n+m)!}}\,
  P_n^{m}(\cos\theta)\,\Exp{im\phi}
\end{equation}
via
\begin{subequations}
  \label{eqs:Y:Ups:The}
  \begin{eqnarray}
    \label{eq:Y:The}
    \Theta_n^m(\vec{r}) &=& \big[(n-m)!\,(n+m)!\big]^{1/2}\,r^{-n-1}\,
      Y_n^m(\hat{\vec{r}}),
    \\[0.5ex]
    \label{eq:Y:Ups}
    \Upsilon_n^m(\vec{r}) &=&\big [(n-m)!\,(n+m)!\big]^{-1/2}\,r^n\,
      Y_n^m(\hat{\vec{r}}).
  \end{eqnarray}
\end{subequations}
Table~\ref{tab:The} gives the first few harmonics in terms of $x,y,z$.

\subsection{Spherical-harmonic FMM}
\label{app:spherical}
In order to derive the relations for spherical-harmonic FMM, one must obtain the
equivalent to the Cartesian Taylor expansion (\ref{eq:fmm:greens}) and shift
operations (\ref{eq:M:shift},e). Via induction one can show that when applied to
harmonic functions
\begin{equation} \label{eq:dot:nabla}
  \frac{1}{n!}(\vec{x}\cdot\vec{\nabla})^n =
  \sum_{m=-n}^{n} \Upsilon_n^{m\ast}(-\vec{x})\,\Delta_n^m =
  \sum_{m=-n}^{n} \Upsilon_n^{m}(-\vec{x})\,\Delta_n^{m\ast},
\end{equation}
which gives the translation operator for harmonic functions
\begin{equation} \label{eq:shift}
  \exp({\vec{x}\cdot\vec{\nabla}})
  = \sum_{n=0}^\infty \sum_{m=-n}^n \,\Upsilon_n^m(-\vec{x})\,\Delta_n^{m\ast}
  = \sum_{n=0}^\infty \sum_{m=-n}^n \,\Upsilon_n^{m\ast}(-\vec{x})\,\Delta_n^{m}.
\end{equation}
When applying this to the harmonic Greens function, one gets
\begin{equation} \label{eq:Sph:r-x}
  \frac{1}{|\vec{r}-\vec{x}|} = \sum_{n=0}^\infty \sum_{m=-n}^{n}
  \Upsilon_n^{m}(\vec{x})\,\Theta_n^{m\ast}(\vec{r}),
\end{equation}
which, because of equations (\ref{eqs:Y:Ups:The}), is equivalent to the standard
form (\ref{eq:spherical}) and converges for $r>x$. Translating once again and
employing (\ref{eq:Delta:prod}) yields
\begin{equation} \label{eq:Sph:r-x-y}
  \frac{1}{|\vec{r}-\vec{x}-\vec{y}|} =
  \sum_{k=0}^\infty\;\sum_{l=-k}^{k}\;\sum_{n=0}^\infty \sum_{m=-n}^{n}
  \Upsilon_n^{m}(\vec{x})\,\Upsilon_k^{l}(\vec{y})\,
  \Theta_{n+k}^{m+l\ast}(\vec{r}),
\end{equation}
which converges for $r>|\vec{x}+\vec{y}|$. Comparing (\ref{eq:Sph:r-x}) and
(\ref{eq:Sph:r-x-y}) one finds immediately the translation formula
\begin{equation} \label{eq:Ups:shift}
  \Upsilon_n^m(\vec{x}+\vec{y}) = \sum_{k=0}^n\sum_{l=-k}^k
  \Upsilon_k^{l}(\vec{x})\,\Upsilon_{n-k}^{m-l}(\vec{y}).
\end{equation}
When applying the translation operator (\ref{eq:shift}) to $\Theta_n^m$,
one gets
\begin{equation} \label{eq:The:shift}
  \Theta_n^m(\vec{x}+\vec{y}) = \sum_{k=0}^\infty\sum_{l=-k}^k
  \,\Upsilon_k^{l\ast}(-\vec{y})\,\Theta_{n+k}^{m+l}(\vec{x}).
\end{equation}
As the Cartesian FMM relations (\ref{eqs:fmm:cartesian}) were based on equation
(\ref{eq:fmm:greens}), the spherical harmonic FMM relations (\ref{eqs:fmm}) are
based on equation (\ref{eq:Sph:r-x-y}), which for $\psi=|\vec{r}|^{-1}$ is
completely equivalent but computationally more efficient.

\begin{table*}
  \caption{The real-valued functions
    $(n-m)!(n+m)!U_n^m(\vec{r})=r^{2n+1}T_n^m(\vec{r})$ for $n\le6$. See
    equations~(\ref{eqs:Unm:Tnm}) for the relations to 
    $\Upsilon_n^m(\vec{r})$ and $\Theta_n^m(\vec{r})$. \label{tab:The}}
  \begin{center}
    \tabcolsep4pt
    \begin{tabular}{@{}r||l|l|r@{\,}l|r@{\,}l|r@{\,}l|r@{\,}l|r@{\,}l@{}}
      $m\backslash n$ &0& 1& 
      \multicolumn{2}{c}{2} &
      \multicolumn{2}{c}{3} &
      \multicolumn{2}{c}{4} &
      \multicolumn{2}{c}{5} &
      \multicolumn{2}{c}{6}
      \\ \hline &&&&&&&&&&&&
      \\[-2ex]
      6 & & & && && && &&
      10395&($x^6{-}$15$x^4y^2{+}$15$x^2y^4{-}y^6$)
      \\[.3ex]
      5 & & & && && &&
      945   & $x$($x^4{-}$10$x^2y^2{+}$5$y^4$) &
      10395 & $xz$($x^4{-}$10$x^2y^2{+}$5$y^4$)
      \\[.3ex]
      4 & & & && &&
      105 & ($x^4{-}$6$x^2y^2{+}y^4$) &
      945 & $z$($x^4{-}$6$x^2y^2{+}y^4$) &
      945 & ($x^4{-}$6$x^2y^2{+}y^4$)(11$z^2{-}r^2$)
      \\[.3ex]
      3 & & & &&
      15  & $x$($x^2{-}$3$y^2$) &
      105 & $xz$($x^2{-}$3$y^2$) &
      105 & $x$($x^2{-}$3$y^2$)(9$z^2-r^2$) &
      945 & $zx$($x^2{-}$3$y^2$)(11$z^2{-}$3$r^2$)
      \\[.3ex]
      2 & & &
      3   & ($x^2{-}y^2$) &
      15  & $z$($x^2{-}y^2$) &
      15  & ($x^2{-}y^2$)(7$z^2{-}r^2$) &
      315 & $z$($x^2{-}y^2$)(3$z^2{-}r^2$) &
      315 & ($x^2{-}y^2$)(33$z^4{-}$18$z^2r^2{+}r^4$)
      \\[.3ex]
      1 & & $x$ &
      3   & $xz$ &
      3   & $x$(5$z^2{-}r^2$) &
      15  & $xz$(7$z^2{-}$3$r^2$) &
      45  & $x$(21$z^4{-}$14$z^2r^2{+}r^4$)&
      315 & $xz$(33$z^4{-}$30$z^2r^2{+}$5$r^4$)
      \\[.3ex]
      0 & 1 & $z$ &
      3  & $z^2{-}r^2$ &
      3  & $z$(5$z^2{-}$3$r^2$) &
      3  & (35$z^4{-}$30$z^2r^2{+}$3$r^4$) &
      15 & $z$(63$z^4{-}$70$z^2r^2{+}$15$r^4$) &
      45 & (231$z^6{-}$315$z^4r^2{+}$105$z^2r^4{-}$5$r^6$)
      \\[.3ex]
      ${-}$1 & & $y$ &
      3   & $yz$ &
      3   & $y$(5$z^2{-}r^2$) &
      15  & $yz$(7$z^2{-}$3$r^2$)&
      45  & $y$(21$z^4{-}$14$z^2r^2{+}r^4$)&
      315 & $yz$(33$z^4{-}$30$z^2r^2{+}$5$r^4$)
      \\[.3ex]
      ${-}$2 & & &
      6   & $xy$ &
      30  & $xyz$ &
      30  & $xy$(7$z^2{-}r^2$) &
      630 & $xyz$(3$z^2{-}r^2$) &
      630 & $xy$(33$z^4{-}$18$z^2r^2{+}r^4$)
      \\[.3ex]
      ${-}$3 & & & &&
      15  & $y$(3$x^2{-}y^2$) &
      105 & $yz$(3$x^2{-}y^2$) &
      105 & $y$(3$x^2{-}y^2$)(9$z^2-r^2$) &
      945 & $zy$(3$x^2{-}y^2$)(11$z^2{-}$3$r^2$)
      \\[.3ex]
      ${-}$4 & & & && &&
      420  & $xy$($x^2{-}y^2$) &
      3780 & $xyz$($x^2{-}y^2$) &
      3780 & $xy$($x^2{-}y^2$)(11$z^2{-}r^2$)
      \\[.3ex]
      ${-}$5 & & & && && &&
      945   & $y$(5$x^4{-}$10$x^2y^2{+}y^4$) &
      10395 & $yz$(5$x^4{-}$10$x^2y^2{+}y^4$)
      \\[.3ex]
      ${-}$6 & & & && && && &&
      20790 & $xy$(3$x^2{-}y^2$)($x^2{-}$3$y^2$)
    \end{tabular}
  \end{center}
\end{table*}

\subsection{Implementation details}
\label{app:spherical:details}
\subsubsection{Recursive evaluation of spherical harmonics}
\label{app:spherical:recurse}
One may also obtain the relations
\begin{eqnarray}
  \label{eq:dTnm}
  \Delta_k^l\,\Theta_n^m(\vec{x}) &=& \Theta_{n+k}^{m+l}(\vec{x}), \\
  \label{eq:dUnm}
  \Delta_k^l\,\Upsilon_n^m(\vec{x}) &=& (-1)^{k+l}\,\Upsilon_{n-k}^{m+l}(\vec{x}).
\end{eqnarray}
The first one follows immediately from equations~(\ref{eq:Delta:prod}) and
(\ref{eq:The:def}), while the second can be deduced by equating
(\ref{eq:Ups:shift}) to $\Upsilon_n^m(\vec{x}+\vec{y})$ obtained by applying the
translation operator (\ref{eq:shift}). From these two relations combined with
the operator relation (\ref{eq:Delta:prod}) and the definitions
(\ref{eq:The:def}) and (\ref{eq:Ups:def}), one can obtain numerous recurrence
relations. For example, (omitting the arguments for brevity)
\begin{eqnarray}
  r^2\,\Theta_n^m  &=& (2n-1)\,z\,\Theta_{n-1}^m
  - ([n-1]^2-m^2)\,\Theta_{n-2}^m, \\
  \label{eq:Ups:rec}
  (n^2-m^2)\Upsilon_n^m &=& (2n-1)\,z\,\Upsilon_{n-1}^m
  - r^2\,\Upsilon_{n-2}^m,
\end{eqnarray}
which are equivalent to the recurrence relation
\citep[][eq.~8.733.2]{GradshteynRyzhik1994} for associated Legendre functions
and, together with
\begin{equation} \label{eq:The:nn}
  \Theta_n^n = (2n-1)\,\frac{x+iy}{r^2}\,\Theta_{n-1}^{n-1}
  \qquad\text{and}\qquad
  \Upsilon_n^n = \frac{x+iy}{2n}\,\Upsilon_{n-1}^{n-1}
\end{equation}
as well as their counterparts for $m=-n$, allow for an efficient and stable
evaluation of $\Theta_n^m(\vec{r})$ and $\Upsilon_n^m(\vec{r})$.

Differentiating these relations with respect to time, one obtains recursion
relations for the time derivatives of the harmonic functions. For example,
\begin{equation}
  (n^2-m^2)\dot{\Upsilon}_n^m
    = (2n-1)\left[\dot{z}\Upsilon_{n-1}^m+ z\,\dot{\Upsilon}_{n-1}^m\right]
    - 2\vec{r}{\cdot}\dot{\vec{r}}\,\Upsilon_{n-2}^m-r^2\dot{\Upsilon}_{n-2}^m.
\end{equation}
Alternatively, from equations~(\ref{eq:dTnm}) and (\ref{eq:dUnm}) one may also
directly derive
\begin{eqnarray}
  \dot{\Theta}_n^m &=&
  \dot{\xi}\,\Theta_{n+1}^{m+1} -
  \dot{z}\,\Theta_{n+1}^{m} +
  \dot{\eta}\,\Theta_{n+1}^{m-1},\quad\text{and}\\[0.5ex]
  \dot{\Upsilon}_n^m &=&
  \dot{\xi}\,\Upsilon_{n-1}^{m+1} +
  \dot{z}\,\Upsilon_{n-1}^{m} +
  \dot{\eta}\,\Upsilon_{n-1}^{m-1}.
\end{eqnarray}

\subsubsection{Real-valued spherical harmonics}
\label{app:spherical:real}
Because of the anti-symmetry relation (\ref{eq:sym}), the complex spherical
harmonics defined above are redundant: there are only $2n+1$ independent (real)
harmonics per order, in agreement with the counting in
Section~\ref{app:Cartesian}. Hence, for any practical application one needs an
appropriately reduced set of $2n+1$ real-valued independent spherical harmonics
per order. The simplest option is to consider real and imaginary parts of the
complex-valued harmonics with $m\ge0$:
\begin{subequations} \label{eqs:Unm:Tnm}
  \begin{equation} \label{eq:Unm}
    U_n^m \equiv \left\{\begin{array}{l@{\;=\;}ll}
    \Im(\Upsilon_n^{|m|}) &
    \frac{1}{2i} (\Upsilon_n^{-m}-\Upsilon_n^{-m\ast}) \qquad & m<0, \\[1ex]
    \Re(\Upsilon_n^{|m|}) &
    \frac{1}{2 } (\Upsilon_n^{ m}+\Upsilon_n^{ m\ast}) \qquad & m\ge 0;
    \end{array}\right.
  \end{equation}
  and
  \begin{equation} \label{eq:Tnm}
    T_n^m \equiv \left\{\begin{array}{l@{\;=\;}ll}
    \Im(\Theta_n^{|m|}) &
    \frac{1}{2i} (\Theta_n^{-m}-\Theta_n^{-m\ast}) \qquad & m<0, \\[1ex]
    \Re(\Theta_n^{|m|}) &
    \frac{1}{2 } (\Theta_n^{ m}+\Theta_n^{ m\ast}) \qquad & m\ge 0.
    \end{array}\right.
  \end{equation}
\end{subequations}
The relevant relations for these real-valued spherical harmonics are best
directly transcribed from the corresponding complex relations.

\subsection{Accelerating FMM relations}
\label{app:spherical:fast}
The FMM kernels M2L, M2M, and L2L (equations~\ref{eq:M2L},d,e) all require
$\mathcal{O}(p^4)$ operations. However, if the interactions or translations
are along the $z$-axis, the costs are only $\mathcal{O}(p^3)$ because
$\Upsilon_n^m(\hat{\vec{z}})= \delta_{m0}/n!$.

One method to exploit this is to first translate along the $z$-axis and then
perpendicular to the $z$-axis. For a vector $\vec{r}_\perp$ perpendicular to the
$z$-axis, $\Upsilon_n^m(\vec{r}_\perp)$ vanishes whenever $n+m$ is even. This
implies that a translation along $\vec{r}_\perp$ can be done faster than a
general translation (in the limit of $p\to\infty$, twice as fast).

This splitting method cannot be applied to the M2L kernel~(\ref{eq:M2L})
(because it is not a translation), which occurs many more times in the FMM
algorithm than the M2M and L2L kernels. To accelerate the M2L kernel, one can
exploit that a rotation only costs $\mathcal{O}(p^3)$ operations, too. Thus, if
one first rotates into a frame in which the interaction is along the $z$ axis,
applies the M2L kernel in the rotated frame, and finally rotates back into the
original frame, the total costs are still $\mathcal{O}(p^3)$.

\subsubsection{Fast rotations} \label{app:rot}
Since the spherical harmonics are homogeneous, a rotation (as opposed to a
translation) does not mix between different orders $n$, and consequently the
operation count is $\mathcal{O}(p^3)$. Thus, a general rotation is of the form
\begin{equation} \label{eq:Ylm:rot}
  Y_n^m(\tilde{\vec{r}}) = \sum_{l=-n}^n \mathsf{\Gamma}_n^{ml}\,Y_n^l(\vec{r}),
\end{equation}
where $\tilde{\vec{r}}$ denotes the vector $\vec{r}$ in the rotated
frame. Unfortunately, the matrices $\tens{\Gamma}_{\!\!n}$, also known as Wigner
functions, are generally dense and non-trivial functions of the Euler
angles. However, a rotation by angle $\alpha$ around the $z$ axis is simple:
\begin{equation} \label{eq:Ylm:rot:z}
  Y_n^m(\tilde{\vec{r}}) = \Exp{-im\alpha}\,Y_n^m(\vec{r})
\end{equation}
with an operation count of only $\mathcal{O}(p^2)$. With this one can build a
general rotation by first rotating around the $z$-axis, then swapping $z$ and
$x$, rotating again about the $z$-axis (the $x$-axis of the original frame),
swapping $z$ and $x$ again, and performing a final rotation around the
$z$-axis. Like rotations, swapping coordinate axes does not mix between
different orders $n$ and can be represented as
\begin{equation} \label{eq:Ylm:swap}
  \Theta_n^m  (\tilde{\vec{r}}) = \sum_{l=-n}^n \mathsf{B}_n^{ml}\,
\Theta_n^l(\vec{r}),
\end{equation}
where now $\tilde{\vec{r}}$ denotes the vector $\vec{r}$ in the frame obtained
by swapping two Cartesian coordinates. The important difference between
equations~(\ref{eq:Ylm:rot}) and (\ref{eq:Ylm:swap}) is that the matrices
$\tens{B}_n$ are \emph{constants}. Recursive relations for these swap matrices
can be derived via the operator algebra of
Section~\ref{app:spherical:complex}. For example, for swapping $x$ and $z$, one
finds
\begin{equation}
2\tilde{\Delta}_1^0 = \Delta_1^1-\Delta_1^{-1}
\quad\text{and}\quad
2\tilde{\Delta}_1^{\pm1} = \Delta_1^1+\Delta_1^{-1} \pm 2\Delta_1^0,
\end{equation}
with which one can derive the recurrence relations
\begin{equation}
  \begin{array}{lclclcl}
    2\mathsf{B}_{n+1}^{ml} &=& \mathsf{B}_n^{m\,l-1} &-&\mathsf{B}_n^{m\,l+1},
    \\[1ex]
    2\mathsf{B}_{n+1}^{m\pm1\,l} &=& \mathsf{B}_n^{m\,l-1}&+&\mathsf{B}_n^{m\,l+1}
    &\pm&2\mathsf{B}_n^{m\,l},
  \end{array}
\end{equation}
where it is understood that $\mathsf{B}_n^{m\,l}=0$ for $|l|>n$. A similar
exercise for swapping $y$ and $z$ reveals that the swap matrices are given by
$i^{m-l}\mathsf{B}_n^{ml}$, while the corresponding swap matrices for
$\Upsilon_n^m$ are given by the transpose (because these matrices are
orthonormal and the product~(\ref{eq:Sph:r-x}) is invariant under coordinate
swapping). Whereas the matrices $\tens{B}_n$ are dense, the corresponding
matrices for the real-valued harmonics (equations \ref{eqs:Unm:Tnm}) are not
\citep{PinchonHoggan2007}. For example, the matrices for swapping $x$ and $z$
for $\Theta_4^m$ and $T_4^m$ are (omitting zero entries)
\begin{eqnarray}
  \frac{1}{16} &\times& \left[
  \begin{array}{rrrrrrrrr}
     1& -8& 28&-56& 70&-56& 28& -8&  1 \\
    -1&  6&-14& 14&   &-14& 14& -6&  1 \\
     1& -4&  4&  4&-10&  4&  4& -4&  1 \\
    -1&  2&  2& -6&   &  6& -2& -2&  1 \\
     1&   & -4&   &  6&   & -4&   &  1 \\
    -1& -2&  2&  6&   & -6& -2&  2&  1 \\
     1&  4&  4& -4&-10& -4&  4&  4&  1 \\
    -1& -6&-14&-14&   & 14& 14&  6&  1 \\
     1&  8& 28& 56& 70& 56& 28&  8&  1
  \end{array} \right]
  \qquad\text{and}\\
  \frac{1}{8} &\times& \left[
  \begin{array}{rrrrrrrrr}
     &  \phantom{-}8&   & \phantom{-}56&   &   &   &   &    \\
   \phantom{-}1&   & \phantom{-}14&   &   &   &   &   &    \\
     &  4&   & -4&   &   &   &   &    \\
    1&   & -2&   &   &   &   &   &    \\
     &   &   &   &  3&   & -4&   &  1 \\
     &   &   &   &   & -6&   &  \phantom{-}2&    \\
     &   &   &   & -5&   &  4&   &  1 \\
     &   &   &   &   & \phantom{-}14&   &  6&    \\
     &   &   &   & \phantom{-}35&   & 28&   &  1
  \end{array} \right],
\end{eqnarray}
respectively. Thus, this method of achieving a general rotation not only avoids
the (recursive) computation of the Wigner functions $\tens{\Gamma}_n$ (which
itself costs $\mathcal{O}(p^3)$ operations), but also benefits from the facts
that the swap matrices $\tens{B}_n$ have $\approx4$ times fewer non-zero
entries than the $\tens{\Gamma}_n$ and are known a priori, such that they can
be `hard-wired' into computer code.

\subsubsection{A fast M2L kernel} \label{app:fast:inter}
With these preliminaries, one can finally put together an accelerated
$\mathcal{O}(p^3)$ version for performing the M2L kernel (\ref{eq:M2L}). Let
$(x,y,z)=\vec{r}$, then one first rotates the multipoles $\mathcal{M}_k^l$
(around the $z$-axis) by angle $\alpha_z=\arctan(y/x)$, swaps $x$ and $z$,
rotates by $\alpha_x=\arctan\sqrt{x^2+y^2}/z$, and swaps $x$ and $z$ back. The
obtained $\tilde{\mathcal{M}}_k^l$ has $z$ axis aligned with the interaction
direction, and the M2L kernel can be performed via
\begin{equation} \label{eq:M2L:z}
  \tilde{\mathcal{F}}_n^m(\vec{z}_B) =
  \sum_{k=|m|}^{p-n} (-1)^m\,\tilde{\mathcal{M}}_k^{m}(\vec{z}_A)\,
  \frac{(n+k)!}{r^{n+k+1}}.
\end{equation}
Finally, one must rotate $\tilde{\mathcal{F}}_n^{m}$ back to the original frame
by first swapping $x$ and $z$, rotating by $-\alpha_x$, swapping $x$ and $z$
again, followed by a final rotation by $-\alpha_z$.

These rotations and swaps can be accelerated further by exploiting that in
(\ref{eq:M2L:z}) only multipoles $\tilde{\mathcal{M}}_n^m$ with $|m|\le
\min\{n,p-n\}$ are needed and, similarly, that $\tilde{\mathcal{F}}_n^m=0$ for
$|m|>\min\{n,p-n\}$. As Fig.~\ref{fig:interact} demonstrates, the overhead due
to the rotations pays off already for $p=5$.

\section{The energy error of a simulation}
\label{app:error}
The gravitational forces (and potentials) used in $N$-body simulations always
carry some error. When using direct summation, this is solely due to round-off
errors, while for approximate methods the approximation error should dominate
round-off. Here, I investigate the consequences of these errors for the
non-conservation of the total energy.

\subsection{The energy error due to force errors}
\label{app:error:forces}
Consider, the energy error generated by acceleration errors $\delta\vec{a}_b$
after one time step $\tau$
\begin{equation} \label{eq:dEtot}
  \textstyle
  \delta E_{\mathrm{tot}} = 
  \tau \sum_b\mu_b \, \dot{\vec{x}}_b\cdot\delta\vec{a}_b.
\end{equation}
Because the $\delta\vec{a}_b$ are not correlated with the velocities
$\dot{\vec{x}}_b$, their dot products largely cancel and $\delta
E_{\mathrm{tot}}$ will be small. In order to estimate its amplitude, let us
assume $\tau=\eta\sigma/\bar{a}$ with $\eta\ll1$, velocity dispersion $\sigma$,
and typical acceleration $\bar{a}$. If further assuming virial equilibrium and a
relative acceleration error $\varepsilon$,
\begin{equation}
   |\delta E_{\mathrm{tot}}|/|E_{\mathrm{tot}}| \sim \eta\varepsilon/
   \sqrt{N}.
\end{equation}
Over time this accumulates in the fashion of a random walk and after one
dynamical time or $\sim\eta^{-1}$ time steps
\begin{equation} \label{eq:dEtot:est}
  |\delta E_{\mathrm{tot}}|/|E_{\mathrm{tot}}|
  \sim \varepsilon \sqrt{\eta/N}.
\end{equation}
Thus, the relative energy error resulting from the force errors alone is much
smaller than $\varepsilon$, simply because it is some average over many force
errors.

\subsection{The measurement error}
\label{app:error:measure}
In order to measure the total energy, one must also calculate the individual
particle potentials $\Psi_b$ (which are otherwise not required for the
simulation). Assuming that the $\Psi_b$ are computed with relative error
$\varepsilon$, the resulting error for the total energy is
\begin{equation}
  |\delta E_{\mathrm{tot}}|/|E_{\mathrm{tot}}| \approx 2\varepsilon/\sqrt{N}.
\end{equation}
If the same precision $\varepsilon$ is used for computing the particle
potentials and accelerations, this is much larger than the energy
error~(\ref{eq:dEtot:est}) due to force errors.

\subsection{Approximate gravity solvers}
\label{app:error:approx}
The situation is different for approximative methods, such as the tree code,
FMM, and mesh-based techniques. All of these approximate the true potential, but
use the exact derivatives of the approximated potential for the accelerations.
Therefore, the total \emph{approximated} energy should be conserved (modulo
round-off errors), even if the approximation is poor.

For the FMM and the tree code the situation is actually different, because the
approximated potential is not globally continuous but only piece-wise. This is
because the concrete form of the approximation used for a given particle depends
on its position (which determines how FMM approximates each pair-wise force). A
particle crossing a boundary between such continuous regions suffers a jump in
the (approximated) potential, and hence energy, while the corresponding kick in
velocity (to conserve energy) is ignored. These discontinuities are part of the
approximation error and their amplitudes proportional. The implication is that
for the tree code and FMM energy is not conserved (even for accurate time
integration) and the degree of non-conservation actually reflects the amplitude
of the approximation errors in an average sense.

\label{lastpage}
\end{document}